\documentclass[sort&compress,english,preview,5p]{elsarticle}

\usepackage{lineno,hyperref}
\modulolinenumbers[5]
\usepackage[mode=buildnew]{standalone}

\journal{Ultramicroscopy}

\usepackage{graphicx}
\usepackage{xcolor}
\usepackage{tabularx}
\usepackage{amsmath}
\usepackage{physics}

\begin{document}

\begin{frontmatter}

\title{Guided progressive reconstructive imaging: a new quantization-based framework for low-dose, high-throughput and real-time analytical ptychography}

\author[emat,nano]{Hoelen L. Lalandec Robert\corref{cor1}}
\author[emat,nano]{Arno Annys}
\author[emat,nano]{Tamazouzt Chennit}
\author[emat,nano]{Jo Verbeeck}

\cortext[cor1]{Corresponding author.\\Email: hoelen.lalandecrobert@uantwerpen.be}
\address[emat]{Electron Microscopy for Materials Science (EMAT), University of Antwerp, Groenenborgerlaan 171, 2020 Antwerp, Belgium}
\address[nano]{NANOlight Center of Excellence, University of Antwerp, Groenenborgerlaan 171, 2020 Antwerp, Belgium}

\begin{abstract}
    By profiting from recent developments in detector technologies, making it possible to access a stream of detection events with few-ns time resolutions, a new ptychographic workflow is established. This methodological framework, referred to as guided progressive reconstructive imaging, relies on a quantization-based description of the acquired intensity, through an elementary derivation. Established direct phase retrieval solutions, such as the Wigner distribution deconvolution approach, can then be adapted to a continuous treatment of received counts, with no need for a dense data representation. Consequently, the result is obtained in the form of a progressively improving estimate, while providing immediate user feedback thanks to a processing speed high enough to surpass the acquisition bandwidth. This fast measurement is enabled by the cumulative usage of a pre-calculated library of kernel-limited functions, accumulating count-wise contributions as a function of the triggered detector pixel. Hence, the reconstruction offers the same advantages of direct phase retrieval methods, in particular a high dose-efficiency and the absence of complex convergence dynamics, with much less stringent restrictions on the field of view than is typical in current alternatives. Its implementation is also significantly more straightforward and flexible. Overall, this work constitutes a major evolution in the state-of-the-art, facilitating repeatable and low-dose experiments with high accessibility, and being applicable to electron-based imaging, X-ray diffraction and optical microscopy.
\end{abstract}


\end{frontmatter}

\section*{Introduction}
    
    Ptychography \cite{Rodenburg2019,Wang2025a}, originally prompted by ref. \cite{Hoppe1969,Hoppe1969a,Hoppe1969b} and having developed in parallel to coherent diffractive imaging methods \cite{Miao2025}, constitutes a powerful approach to computational microscopy, widely applicable across scattering experiments. Specifically, it consists in the treatment of a multidimensional intensity acquisition, aiming to recover a map of the phase/amplitude modulations imposed by a solid-state specimen to the incident wave. In its dominant form, this technique relies on a convergent illumination, i.e. a focused probe, scanned over the object within a two-dimensional raster \cite{McCallum1992,Faulkner2004,Rodenburg2004,Thibault2008,Guizar-Sicairos2008}. In parallel, the detection of scattered wavefronts, in the form of convergent beam electron diffraction (CBED) patterns, is performed in the far-field.
    
    This experimental setup is nowadays available in scanning transmission electron microscopy (STEM) \cite{Muller2012} through the usage of a direct electron detector (DED) \cite{McMullan2007,Milazzo2010,Ballabriga2011,Poikela2014,Ryll2016,Tate2016,Llopart2022,Zambon2023,Ercius2024}, in a recording geometry usually referred to as 4D-STEM \cite{Yang2015a} or momentum-resolved STEM (MR-STEM) \cite{Muller-Caspary2018a}. Furthermore, capacities for short acquisitions, combined with specialized softwares, have permitted an exploration of real-time imaging in MR-STEM \cite{Strauch2021,Yu2022a,Bangun2023,Weber2024}, i.e. the continuous generation of an interpretable micrograph in-acquisition.
    
    Within the last few years, this growing field has also profited from novel event-driven detectors (EDD) based on the Timepix3 \cite{Poikela2014,Frojdh2015} and, more recently, the Timepix4 \cite{Llopart2022} technologies. In particular, they allow the continuous generation of a stream of detection event coordinates, rather than a collection of dense 2D scattering frames. Hence, this ensures higher data throughput \cite{Auad2024,Kuttruff2024}, only limited by the maximum hit rate of the chip and thus usable in a certain range of sustainable beam currents, appropriate for low-dose measurements. By enabling sub-$\mu$s probe position dwell times \cite{Jannis2022,Annys2025}, this also led to a significant increment in the achievable number of scan points, extending electron ptychography to wider fields of view. This new EDD-based implementation, useful in the X-ray diffraction case \cite{Pfeiffer2018} as well, is thus especially relevant for beam-sensitive specimens \cite{Egerton2019}.
    
    Moreover, for this particular application, two general interests of ptychographic computational imaging are its high dose-efficiency \cite{Pennycook2019,OLeary2020,OLeary2021,Jilek2025,LalandecRobert2025,Dearg2025,Varnavides2025} and its strong sensitivity to light atoms \cite{Yang2016,Yang2017,Wang2017a,Wen2019,Leidl2023}, compared to conventional STEM methods. Those advantages have been experimentally tested for a variety of materials, including Li-rich materials \cite{Lozano2018,Song2022,Song2024c}, zeolites \cite{Sha2023a,Zhang2023,Dong2023,Mitsuishi2023,Li2026b}, polymers \cite{Hao2023}, 2D materials \cite{Wen2019,Chen2024c,Loh2025,Hofer2025}, metal-organic frameworks \cite{Li2025}, perovskites \cite{Scheid2023,Yuan2025b} and biological objects \cite{Zhou2020,Pei2023,Kucukoglu2024}. As a side-note, in the latter case, large image sizes and higher throughput would be particularly useful for single-particle analysis \cite{Cheng2015,Nakane2020,Pei2023} with improved performance.
    
    For the vast majority of situations, this outstanding approach nevertheless remains limited by the necessity of a time-consuming calculation, performed post-acquisition. This is an inherent aspect of the sophisticated iterative optimization methods \cite{Wang2025a} that are often used. Specifically, the large volumes of data involved and the non-linear update pathway imply significant computational costs and memory needs. Ptychography thus typically requires high-cost hardware with specialized parallelization tools \cite{Wakonig2020,Yu2022,Wang2022c,Loetgering2023,Gilgenbach2026,Lee2025,Skoupy2025,Du2025}, and may even warrant support from high-performance computing facilities \cite{Jones2022,Mukherjee2022,Wang2022c,Welborn2024}.
    
    Moreover, difficulties in reaching convergence are common in conditions where the invested dose is low \cite{Katvotnik2013,Dearg2025,Chennit2025}. This last aspect, especially considering that popular algorithms tend to provide distinct and parameter-specific results even under low Poisson noise \cite{Maiden2017,Bangun2022,Leidl2024,Maiden2024,Chennit2025}, questions their reproducibility.
    
    Crucially, non-iterative solutions for focused probe-based phase retrieval have been proposed as well, relying on integrated center of mass (iCoM) \cite{Muller2014,Lazic2016,Yucelen2018,Yu2022a} imaging and analytical ptychography. The latter includes the Wigner distribution deconvolution (WDD) \cite{Bates1989,Rodenburg1992,McCallum1992,Chapman1996,Nellist1994,Li2014} and the sideband integration (SBI) \cite{Rodenburg1993,Pennycook2015,Yang2015b,Yang2016a} methods. Those three approaches thus constitute fast, direct and fully repeatable processing tools, appropriate for real-time measurements \cite{Strauch2021,Yu2022a,Bangun2023}.
    
    Another option is the optimum bright field (OBF) method \cite{Ooe2021,Ooe2023,Ooe2024,Ooe2026} which, like SBI imaging, is founded on the weak phase object approximation (WPOA) \cite{Cowley1972}. OBF-STEM can generally be described as a linear superposition of scan position-shifted kernels, each being associated to a specific, arbitrarily shaped, detector in the Fraunhofer plane. Hence, the convolution of the scanned detector signals with those kernels and their addition lead to an interpretable image. As the method typically relies on a segmented device for differential phase contrast (DPC) \cite{Lohr2012}, it can be seen as an interesting alternative to a naive vector field integration approach \cite{Lazic2016,Yucelen2018}. Moreover, as the workflow is adaptable to the pixels of a DED, it also constitutes a real-space convolutive form of an SBI processing.
    
    In iCoM and SBI, the conversion of 2D scattering frames into a sparse format \cite{Guo2020,Datta2021}, applicable when only a small number of counts are registered \cite{OLeary2020}, was furthermore shown to reduce the computational complexity of a part of the procedure \cite{Pelz2022,Ercius2024}. This naive sparse conversion strategy nevertheless remains very limited, in particular since it does not question the original workflow inherent to a dense representation. As a consequence, state-of-the-art implementations of direct phase retrieval are currently unable to make full use of an EDD-enabled data stream, while showing restrictions in terms of their field of view, due to a non-linearly increasing numerical complexity.
    
    To offer both an alternative to conventional analytical approaches and a welcome complement to iterative ptychography solutions, which would allow fast reconstructions and the live treatment of detection events \cite{Annys2025}, this publication introduces a novel processing tool, relying on a count-wise description for the iCoM, SBI and WDD methods. The new framework, in the following referred to as guided progressive reconstructive imaging (GPRI) \cite{LalandecRobert2025a}, fully leverages quantization in a set of CBED patterns, through an individual, and elementary, processing. This translates into an algorithm consisting in a fully summative procedure, and only requiring the pre-calculation of a library having no relation to the specimen and compiling pixel-specific reconstruction kernels, thus showing some similarity with OBF-STEM.
    
    Importantly, the computational cost of GPRI increases linearly with the amount of detected events. Combined with an specialized processing pipeline, this leads to a drastic reduction in reconstruction times and less restrictions with regards to image size. Moreover, because of the simplicity of its individual treatment steps, real-time imaging can be performed over a wide EDD bandwidth \cite{Annys2025}, with a straightforward generalization to the formation and usage of 2D frames.
    
    In the rest of this publication, the theory justifying the GPRI framework is presented in detail, as well as its implications for the processing of MR-STEM data. Some aspects of numerical implementation are then reviewed, followed by demonstrations based on dose-limited simulations as well as through experimental results. Finally, this publication goes through the interests of this new approach in the wider field, through an extensive discussion.

\section{Importance of acquisition sparsity in focused probe-based computational imaging}
    \label{sec:theorysection1}
    
    \subsection{Limits of state-of-the-art data partitioning strategies}
    \vspace{3pt}
        \label{subsec:datapartitioning}
        
        As was hinted in the introduction, a ptychographic treatment of diffraction patterns presents one fundamental limitation. That is, all the counts received at a single scan position $\vec{r}_s$ are treated as a whole, being part of a single two-dimensional frame resolved along the scattering vector $\vec{q}$. This frame is then used to calculate a spatially localized update to the result, iterated or not.
        
        In conventional analytical ptychography workflows, and excluding the OBF-STEM method \cite{Ooe2021,Ooe2023,Ooe2024,Ooe2026}, the processing furthermore relies on a collective treatment of the full dataset, i.e. involving a $\vec{r}_s$-wise transformation to reach the recoverable specimen frequencies $\vec{Q}$, based on a fast Fourier transform (FFT) algorithm. In that manner, the reconstruction grid is fixed by the scan grid, thus presenting a major restriction in the obtained SBI and WDD results.
        
        To alleviate this latter issue, a scan-frequency partitioning algorithm (SFPA) \cite{LalandecRobert2025} was recently introduced, performing this $\vec{r}_s$-to-$\vec{Q}$ translation step through a discrete Fourier transform (DFT) process. This then allows an explicit separation of each introduced $\vec{r}_s\,/\,\vec{Q}$ couple, resulting in an extreme reduction in memory footprint for the complete workflow, thus more adapted to GPU-based calculation. In particular, processing steps are flexibly segmented among intersections of scan position packets $P_{\vec{r}_s}$ and frequency domains $D_{\vec{Q}}$, both with user-defined sizes. Most interestingly, this solution allows the numerical decoupling of the reconstruction grid from the scan dimensions, i.e. with the $\vec{Q}$-coordinates being defined arbitrarily. This entails a relaxation of conventional sampling conditions, and thus makes it possible to use a defocused electron probe with a sparse scan raster \cite{Hue2010,Song2019}. In prior implementations, this was considered as a strict limitation \cite{Li2014,Pennycook2015}.
        
        On the other hand, under a simple physics viewpoint, the finest possible partitioning of the scattering data is not at the level of the scan positions but is rather due to quantization. Specifically, the incident electrons are received one-by-one by the detector through the deposition of individual, and localized, packets of charges. This remains true regardless of the employed read-out process \cite{Ballabriga2011,Ryll2016}, with or without clustering issues \cite{Mir2017,Kuttruff2024} and independently of the resulting data format or available dynamic range \cite{Tate2016,Cao2018,Philipp2022}.
        
        An application of this concept is the dense-to-sparse data conversion mentioned in the introduction, which permits to perform the initial $\vec{r}_s$-to-$\vec{Q}$ reformulation through a count-by-count DFT \cite{Pelz2022,Ercius2024}. This then alters the regime of numerical complexity, going from a straightforward relation with the scan grid to a finer, and often more advantageous, dependence on the total number of counts received.
        
        This approach nevertheless presents two significant limitations. First, the time taken by count extraction was reported to remain in a range of a few minutes for a scan grid of 1024 by 1024 positions, even with the assistance of a supercomputing facility and for doses in the order of 10$^{3}$ $e^-/\text{\AA}^2$. This prevents live imaging, while limiting comparable applications to high-dose data. Second, since the employed SBI algorithm is still formally identical to the conventional one beyond the initial Fourier transform step, the discrete representation of electron counts remains underexploited and some of the same fundamental limitations are met. Those aspects are discussed in more details in subsection \ref{subsec:numericalaspects}.
    
    \subsection{Information carried by an isolated detection event}
    \vspace{3pt}
        \label{subsec:infofromsinglecounts}
        
        In the rest of this publication, the function $\mathcal{F}$ will represent a Fourier transform, and $\mathcal{F}^{-1}$ its inverse. The convention followed will be
        \begin{equation}
        \begin{split}
            \tilde{\varrho}\left(\vec{v}\right) \, = \, \mathcal{F}\left[\,\varrho\left(\vec{u}\right)\,\right]\left(\vec{v}\right) \, & = \, \sum\limits_{\vec{u}} \, e^{-i 2\pi \vec{v}\cdot\vec{u}} \, \varrho\left(\vec{u}\right) \\
            \varrho\left(\vec{u}\right) \, = \, \mathcal{F}^{-1}\left[\,\tilde{\varrho}\left(\vec{v}\right)\,\right]\left(\vec{u}\right) \, & = \, \sum\limits_{\vec{v}} \, e^{i 2\pi \vec{v}\cdot\vec{u}} \, \tilde{\varrho}\left(\vec{v}\right) \quad,
        \end{split}
        \end{equation}
        with arbitrary normalization. For simplification, integrals will all be written in a discrete form.
        
        In a first step toward devising a reconstruction strategy based on isolated electron incidences, the MR-STEM setup may be first described with a simple multiplicative interaction model. In particular, the accessible scattered intensity $I_{\vec{r}_s}\left(\vec{q}\right)$ results from the far-field propagation of a scan position-wise exit wave $\Psi_{\vec{r}_s}\left(\vec{r}_0\right)$, with $\vec{r}_0$ a coordinate in the specimen plane. This is formulated by
        \begin{equation}
            \label{eq:scatteredintensity1}
            I_{\vec{r}_s}\left(\vec{q}\right) \, = \, \mid \, \mathcal{F}\left[ \, \Psi_{\vec{r}_s}\left(\vec{r}_0\right) \, \right]\left(\vec{q}\right) \, \mid^2 \quad,
        \end{equation}
        and
        \begin{equation}
            \label{eq:scatteredintensity2}
            \Psi_{\vec{r}_s}\left(\vec{r}_0\right) \, = \, P\left( \vec{r}_0 - \vec{r}_s \right) \, T\left(\vec{r}_0\right) \quad,
        \end{equation}
        where $T\left(\vec{r}_0\right)$ is the transmission function representing the specimen and $P\left(\vec{r}_0\right)$ is the electron probe, i.e. a convergent and aperture-limited illumination focused to a small area.
        
        As is shown by equations \ref{eq:scatteredintensity1} and \ref{eq:scatteredintensity2}, the distribution $I_{\vec{r}_s}\left(\vec{q}\right)$ constitutes a two-dimensional map of the weightings associated to each of the $\vec{q}$-tilted plane waves making up $\Psi_{\vec{r}_s}\left(\vec{r}_0\right)$, at the specific probe location $\vec{r}_s$. Under this simple understanding, it becomes possible to envision what information a single detection event, at an arbitrary four-dimensional coordinate $\left[{\vec{r}_s}^{\,\,\,c}\,;\,{\vec{q}}^{\,\,c}\right]$, may provide in a phase retrieval process. Specifically, this individual realization of $I_{\vec{r}_s}\left(\vec{q}\right)$ is a clear proof of the existence of a spatial frequency ${\vec{q}}^{\,\,c}$ in the composition of $\Psi_{{\vec{r}_s}^{\,\,\,c}}\left(\vec{r}_0\right)$. In that sense, an isolated collapse of the far-field-propagated wavefront already represents a conclusive scattering experiment. Intuitively, it should be possible to translate this acquired knowledge, no matter how limited, into a partial, spatially dependent, reconstruction of the object.
    
    \subsection{Proposal for a cumulative phase retrieval method}
    \vspace{3pt}
        \label{subsec:GPRIproposal}
        
        \begin{figure}
            \centering
            \includegraphics[width=0.9\columnwidth]{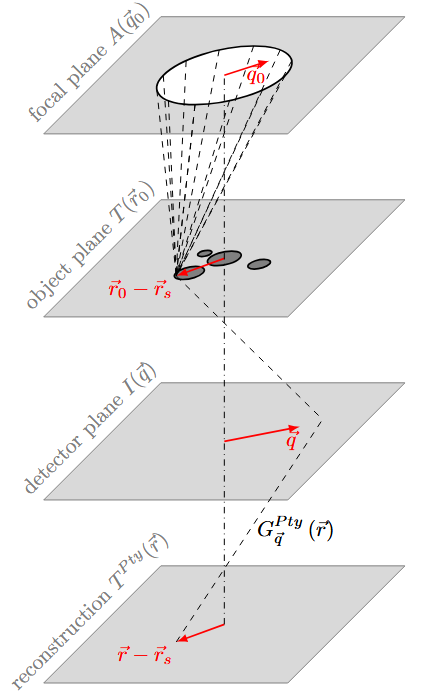}
            \caption{Illustration of a STEM-based scattering experiment and its usage within a basic GPRI process. An electron wave travels from the focal plane $\vec{q}_0$ to the specimen plane $\vec{r}_0$, and propagates to the far-field $\vec{q}$ before collapsing. The resulting detection event leads to a specific contribution to the reconstructed specimen. It is obtained by selection of $G^{Pty}_{\vec{q}}\left(\vec{r}\right)$ within a library, and addition around the current scan position $\vec{r}_s$.}
            \label{fig:basicSTEMGPRIscheme}
        \end{figure}
        
        The reasoning initiated in subsection \ref{subsec:infofromsinglecounts} can be pursued by considering a measurement $T^{Pty}\left(\vec{r}\right)$ of the unknown transmission function $T\left(\vec{r}_0\right)$, retrieved from the full scattering data, with $\vec{r}$ an arbitrary reconstruction space coordinate. Under an analytical phase retrieval approach, i.e. involving a non-iterative processing, a detection event at $\left[{\vec{r}_s}^{\,\,\,c}\,;\,{\vec{q}}^{\,\,c}\right]$ would result into one individual contribution $\Delta T^{Pty}_{\left[{\vec{r}_s}^{\,\,\,c}\,;\,{\vec{q}}^{\,\,c}\right]}\left(\vec{r}\right)$ to this complete measurement. This condition of linearity is already fulfilled with regards to the calculation of an electromagnetic Aharonov-Bohm phase shift \cite{Aharonov1959} in the SBI \cite{Rodenburg1993} and iCoM \cite{Muller2014,Lazic2016} approaches. In the WDD solution \cite{Rodenburg1992}, while phase estimation itself is not linear, as will be clarified in section \ref{sec:theorysection2}, the retrieval of $T\left(\vec{r}_0\right)$ is.
        
        Most importantly, an additive cumulation of the contributions then leads to the same result as a collective treatment, i.e. based on an equivalent multi-count intensity. This is formalized by
        \begin{equation}
            \label{eq:primitiveGPRI1}
            T^{Pty}\left(\vec{r}\right) \, \propto \, \sum\limits_{\left[{\vec{r}_s}^{\,\,\,c}\,;\,{\vec{q}}^{\,\,c}\right]} \, \Delta T^{Pty}_{\left[{\vec{r}_s}^{\,\,\,c}\,;\,{\vec{q}}^{\,\,c}\right]}\left(\vec{r}\right) \quad,
        \end{equation}
        which already implies the possibility for an event-wise treatment of the scattering dataset, maybe even in-acquisition. Note that, in equation \ref{eq:primitiveGPRI1}, the use of a proportionality relation is justified by the need to normalize the intensity afterward. This is clarified in the following. As an additional consequence, the successive calculations and additions of distinct $\Delta T^{Pty}_{\left[{\vec{r}_s}^{\,\,\,c}\,;\,{\vec{q}}^{\,\,c}\right]}\left(\vec{r}\right)$ components should be independent of one another. Similarly, electrons received at the same detector coordinate, and thus for scattering vectors $\vec{q}$ considered equal under pixel size accuracy, lead to identical specimen updates.
        
        Moreover, in analogy to the conventional approach, all contributions constitute an inversion of a single count dataset, i.e. a separate phase space-wise deconvolution by the Wigner distribution of $P\left(\vec{r}_0\right)$ \cite{Bates1989,Rodenburg1992}. That statement alone already implies that the updates should be determined only by the illumination conditions. Therefore, it is the assumed probe model, in combination with the receiving ${\vec{q}}^{\,\,c}$, that should lead to a unique formulation for $\Delta T^{Pty}_{\left[{\vec{r}_s}^{\,\,\,c}\,;\,{\vec{q}}^{\,\,c}\right]}\left(\vec{r}\right)$, known pre-reconstruction. As such, its lateral extent over $\vec{r}$ will reflect the area illuminated in a single acquisition, which already implies a possibility of spatial restriction within a kernel.
        
        Finally, the independence from the underlying $T\left(\vec{r}_0\right)$ and the consistency of $P\left(\vec{r}_0\right)$ throughout the scan imply an equivalence in how each probe position should be treated. Contributions belonging to different locations in the scan grid should thus only differ by a lateral shift.
        
        From those few simple statements, it can be established that
        \begin{equation}
            \label{eq:primitiveGPRI2}
            \Delta T^{Pty}_{[\,\vec{r}_s\,;\,\vec{q}\,]}\left(\vec{r}\right) \, \equiv \, G^{Pty}_{\vec{q}}\left( \vec{r} - \vec{r}_s \right) \quad.
        \end{equation}
        As such, for any event at an arbitrary 4D coordinate $\left[\vec{r}_s\,;\,\vec{q}\right]$, the associated $\Delta_{\left[\vec{r}_s\,;\,\vec{q}\right]} \, T^{Pty}\left(\vec{r}\right)$ may be recovered from a library $G^{Pty}_{\vec{q}}\left(\vec{r}\right)$ of two-dimensional, laterally limited, distributions. In the rest of this publication, they will be referred to as guide functions. In essence, a specific function is pre-calculated for each scattering vector $\vec{q}$ that may be activated in the experiment, while the scan position information only translates into a shift around the concerned $\vec{r}_s$, within the reconstruction window $\vec{r}$. Coming back to the arguments given in subsection \ref{subsec:infofromsinglecounts}, this library then constitutes a tool to translate the knowledge on the scattering vector components of $\Psi_{\vec{r}_s}\left(\vec{r}_0\right)$ into new information on the illuminated object. The resulting interplay of wave propagation, collapse and contribution recovery is illustrated in fig. \ref{fig:basicSTEMGPRIscheme}.
        
        The concepts outlined by equations \ref{eq:primitiveGPRI1} and \ref{eq:primitiveGPRI2} constitute the basis of the guided progressive reconstructive imaging framework introduced in that publication. That is, it is progressive in its workflow, since single counts are treated flexibly and cumulatively, and it is guided, due to being based on known possible contributions for different detector pixels. With this being established, it remains to determine the exact formulation of $G^{Pty}_{\vec{q}}\left(\vec{r}\right)$, and how it relates to conventional methods. This will be done in section \ref{sec:theorysection2}, for WDD ptychography.
    
    \subsection{Role of counting statistics under a restricted dose}
    \vspace{3pt}
        \label{subsec:doseandsparsity}
        
        To better understand how the GPRI procedure relates to the conventional analytical ptychography case, it is relevant to review the consequences of count sparsity in ptychography, and in MR-STEM in general. As explained above, the foundation of this novel method is a discretized representation of the data. Although it may appear trivial at first sight, this representation is rarely employed explicitly.
        
        Typically, in the literature on computational microscopy, acquired diffraction patterns or optical images are treated as being obfuscated by a supplementary Poisson noise \cite{Cederquist1987,Luczka1991} component, leading to a well-defined dose-dependent variance among independent measurements. As such, not only is the phase of the collapsing wavefront not accessible \cite{Drenth1975}, but its amplitude can only be estimated with a certain known precision \cite{Cederquist1987}. This then entails a propagation of noise from detector space to the result. In the case of direct methods, this was investigated empirically in ref. \cite{Pennycook2019,OLeary2020,OLeary2021,Jilek2025,LalandecRobert2025,Dearg2025,Varnavides2025} and, for the iterative variants, in ref. \cite{Godard2012,Katvotnik2013,DAlfonso2016,Leidl2024,Jilek2025,Dearg2025,Chennit2025}. Moreover, when the electron dose is sufficiently low, the acquisition ends up showing clear sparsity \cite{OLeary2020}, i.e. only a few pixels of the electron detector are activated.
        
        In a DED, single electrons may furthermore be counted more than once, mostly due to the travel of generated charge carriers among neighboring pixels. This leads to the formation of clusters having distinct shapes and sizes \cite{Mir2017,VanSchayck2020,Jannis2022,Kuttruff2024}. In a first naive approach to account for this event-wise point-spread effect, a modulation transfer function (MTF) $M\left(\vec{r}_d\right)$ can be introduced. Then, rather than the underlying $I_{\vec{r}_s}\left(\vec{q}\right)$, the intensity $I^{det}_{\vec{r}_s}\left(\vec{q}_d\right)$ that is truly accessible by the detector is given by
        \begin{equation}
            \label{eq:modulationtransferfunction}
            I^{det}_{\vec{r}_s}\left(\vec{q}_d\right) \, = \, \mathcal{F}\left[ \, M\left(\vec{r}_d\right) \,\mathcal{F}^{-1}\left[ \, I_{\vec{r}_s}\left(\vec{q}\right) \, \right]\left(\vec{r}_d\right) \, \right]\left(\vec{q}_d\right) \quad,
        \end{equation}
        with the coordinate system $\vec{q}_d$ explicitly corresponding to the detector pixels, and $\vec{r}_d$ being its inverse reciprocal space. To lift the ambiguity concerning the anisotropy of clusters, $M\left(\vec{r}_d\right)$ is treated as complex-valued in the rest of this work. That is, while $\tilde{M}\left(\vec{q}_d\right)$ is real, it may not be centrally symmetric. In the strictest sense, each cluster should then be considered to have its own complex-valued MTF response, which in practice cannot be included in an experiment. As a result, equation \ref{eq:modulationtransferfunction} is interpretable as including an average information spread effect.
        
        Finally, within the experimental data $I^{exp}_{\vec{r}_s}\left(\vec{q}_d\right)$, each measured electron constitutes a single random event, following the MTF-affected probability weighting $I^{det}_{\vec{r}_s}\left(\vec{q}_d\right)$. The total number $n\left(\vec{r}_s\right) \, = \, \sum\limits_{\vec{q}_d} \, I^{exp}_{\vec{r}_s}\left(\vec{q}_d\right)$ of counts received at a scan position $\vec{r}_s$ is determined by the Poisson distribution, with expectancy $N_{e^-}$, leading to the statistics described e.g. in ref. \cite{Seki2018}.
        
        In a treatment of this four-dimensional acquisition through iterative ptychography, a theoretically sound approach would be to minimize a Poisson-based loss function \cite{Bian2016,Odstrcil2018,Seifert2023,Leidl2024}. However, as explained in the introduction, this is often not desirable for the lowest dose cases, due to leading to possible convergence issues and long calculation times. On the other hand, for the application of an analytical solution, the collection of counts $I^{exp}_{\vec{r}_s}\left(\vec{q}_d\right)$, including a normalization step, can simply be considered to represent the best available estimate of the targeted $I^{det}_{\vec{r}_s}\left(\vec{q}_d\right)$. That is, the experimental data is inserted in the direct processing pipeline as it is.

\section{Establishment of a quantization-based description of analytical ptychography}
    \label{sec:theorysection2}
    
    \subsection{Coherent and elastic electron-specimen interaction under the phase object approximation}
    \vspace{3pt}
        
        In equation \ref{eq:scatteredintensity2}, a simple multiplicative interaction model was introduced. While remaining limited, in particular due to ignoring in-specimen propagation, it leads to a straightforward relationship between the scan position-wise exit wave $\Psi_{\vec{r}_s}\left(\vec{r}_0\right)$, the electron probe $P\left(\vec{r}_0\right)$ and the transmission function $T\left(\vec{r}_0\right)$. In the following, this will be used to establish a formalism for the guide functions mentioned above.
        
        As a first additional step, an assumption can be made on the full coherence of the illumination, leading to
        \begin{equation}
            P\left(\vec{r}_0\right) \, = \, \mathcal{F}^{-1}\left[ \, A\left(\vec{q}_0\right) \, e^{-i\chi\left(\vec{q}_0\right)} \, \right]\left(\vec{r}_0\right) \quad,
        \end{equation}
        where $\chi\left(\vec{q}_0\right)$ is a geometrical aberration function, due to imperfections in the optical system and the defocus of the probe-forming lens, and $A\left(\vec{q}_0\right)$ represents an aperture in the focal plane $\vec{q}_0$. It is given by
        \begin{equation}
            A\left(\vec{q}_0\right) \, = \, 
            \begin{cases}
                \, 1 & \,\,\text{if } \parallel\vec{q}_0\parallel \, < \, q_A \\
                \, 0 & \,\,\text{otherwise}
            \end{cases}
            \quad.
        \end{equation}
        The quantity $q_A\,=\,\sin\left(\alpha\right)\,/\,\lambda$ is the aperture radius in reciprocal space, with $\alpha$ a convergence half-angle and $\lambda$ the relativistically corrected wavelength \cite{Fujiwara1961} of the interacting electrons.
        
        By only treating fully coherent and elastic scattering, usually dominant in thin specimens, it is also possible to use the phase object approximation (POA) \cite{Cowley1972}. Then, the transmission function is expressed by
        \begin{equation}
            T\left(\vec{r}_0\right) \, = \, e^{ i \, \varphi\left(\vec{r}_0\right) } \quad,
        \end{equation}
        thus introducing a phase shift map $\varphi\left(\vec{r}_0\right)$, expected to represent all interaction-induced modifications of the incident wave. In that context, the absence of an amplitude variation in $T\left(\vec{r}_0\right)$ is also reflective of particle conservation, i.e. implying electron-transparency and negligible intensity beyond detector-covered angles. The phase shift is equivalently described as resulting from an electromagnetic, spatially varying, Aharonov-Bohm effect \cite{Aharonov1959}, and follows
        \begin{equation}
            \varphi\left(\vec{r}_0\right) \, = \, \sigma \mu\left(\vec{r}_0\right) \, - \, \frac{2\pi e}{h} \phi_B\left(\vec{r}_0\right) \quad,
        \end{equation}
        where $\mu\left(\vec{r}_0\right)$ is a projected electrostatic potential, i.e. the integral of the three-dimensional potential along the propagation axis, and $\phi_B\left(\vec{r}_0\right)$ is a magnetic flux term, relevant only in concerned materials \cite{Chen2022,Cui2023,Mendoza2025}. The quantity $\sigma$, expressed in $\text{V}^{-1}\cdot\text{m}^{-1}$, is a known interaction parameter \cite{Fujiwara1961}, $h$ is the Planck constant and $e$ is the elementary charge.
    
    \subsection{Conventional phase space-wise workflow}
    \vspace{3pt}
        \label{subsec:phasespacedeconv}
        
        The key concept of analytical ptychography is the conversion of the four-dimensional intensity $I^{det}_{\vec{r}_s}\left(\vec{q}_d\right)$ into a new complex-valued distribution $J_{\vec{Q}}\left(\vec{R}\right)$. This first step constitutes a reformulation of the scattering data within a phase space coordinate system $\left[\vec{Q}\,;\,\vec{R}\right]$, and is ensured through combined Fourier transformations, following
        \begin{equation}
            \label{eq:phasespacereformulation}
            J_{\vec{Q}}\left(\vec{R}\right) \, = \, \sum\limits_{\vec{r}_s} \, \sum\limits_{\vec{q}_d} \,  e^{-i 2\pi \vec{Q}\cdot\vec{r}_s} \, e^{i 2\pi \vec{q}_d\cdot\vec{R}} \, I^{det}_{\vec{r}_s}\left(\vec{q}_d\right) \quad.
        \end{equation}
        Importantly, in an appropriate DFT-based implementation \cite{LalandecRobert2025}, the $\vec{Q}$- and $\vec{R}$-grids are defined numerically and remain flexible. Those spaces are thus decoupled from the original experimental setup and the reformulation can be performed with arbitrary real-space sampling, and while accounting for optical distortions \cite{Robert2021}.
        
        When the ratio of common illuminated area among neighboring scan positions is high enough \cite{Bunk2008}, as can be estimated using a simple two-dimensional cross-correlation metric \cite{Huang2014,LalandecRobert2025}, $J_{\vec{Q}}\left(\vec{R}\right)$ is equal to the product of two Wigner distributions with the MTF, as was shown originally in ref. \cite{Bates1989,Rodenburg1992}. This can be formulated by
        \begin{equation}
            \label{eq:Wignerdistributions1}
            J_{\vec{Q}}\left(\vec{R}\right) \, = \, M\left(\vec{R}\right) \, \Gamma\left(\vec{Q}\,;\,\vec{R}\right) \, \Upsilon\left(\vec{Q}\,;\,\vec{R}\right) \quad,
        \end{equation}
        where $\Upsilon\left(\vec{Q}\,;\,\vec{R}\right)$ only includes information on the specimen, while $\Gamma\left(\vec{Q}\,;\,\vec{R}\right)$ depends on the illumination. Specifically, it is found that
        \begin{equation}
        \begin{split}
            \label{eq:Wignerdistributions2}
            \tilde{\Upsilon}\left(\vec{Q}\,;\,\vec{q}_0\right) \, & = \, \tilde{T}\left(\vec{q}_0\right) \, \tilde{T}^*\left(\vec{q}_0-\vec{Q}\right) \\
            \tilde{\Gamma}\left(\vec{Q}\,;\,\vec{q}_0\right) \, & = \, A\left(\vec{q}_0\right) \, A\left(\vec{q}_0+\vec{Q}\right) \, e^{-i\left( \chi\left(\vec{q}_0\right) - \chi\left(\vec{q}_0+\vec{Q}\right) \right)} \quad,
        \end{split}
        \end{equation}
        Crucially, equations \ref{eq:Wignerdistributions1} and \ref{eq:Wignerdistributions2} provide an opportunity to solve the ptychographic phase problem through a direct deconvolutive treatment of $J_{\vec{Q}}\left(\vec{R}\right)$. Hence, in the conventional WDD paradigm, a measurement $T^{WDD}\left(\vec{r}\right)$ of the specimen transmission function, with diffraction- and coherence-limited resolution \cite{Nellist1994,Oxley2020}, is performed through
        \begin{equation}
            \label{eq:directWDD}
            \tilde{T}\left(\vec{0}\right) \tilde{T}^*\left(-\vec{Q}\right) \, \approx \, \sum\limits_{\vec{R}} \, \frac{ M^*\left(\vec{R}\right) \Gamma^*\left(\vec{Q}\,;\,\vec{R}\right) J_{\vec{Q}}\left(\vec{R}\right) }{ \epsilon \, + \, \mid M\left(\vec{R}\right) \Gamma\left(\vec{Q}\,;\,\vec{R}\right) \mid^2 } \quad,
        \end{equation}
        thus under the form of a Wiener filter. The user-defined parameter $\epsilon$ is a small number introduced there to avoid a numerical divergence at values of $M\left(\vec{R}\right) \Gamma\left(\vec{Q}\,;\,\vec{R}\right)$ below the numerical precision. Beyond that, its only requirement is to be high enough to prevent noise amplification \cite{OLeary2021}, under the available radiation dose.
        
        Following this deconvolution/summation step, the intermediary result needs to be normalized by the square root of its DC component to recover the final reconstructed $T^{WDD}\left(\vec{r}\right)$. A phase shift map $\varphi^{WDD}\left(\vec{r}\right)$ is then obtained by extraction of its angle. In practice, the measurement may also display variations of amplitude, i.e. an empirical absorption potential \cite{Humphreys1968,Cowley1972}. Still assuming a thin specimen, this can typically be related to an insufficient overlap ratio \cite{Bunk2008,Edo2013,Huang2014}, or to coherence losses induced by e.g. inelastic scattering \cite{Mkhoyan2008,Beyer2020,Robert2022,Kwon2026,Mendis2026a} and lattice vibrations \cite{Muller2001,VanDyck2011,VanDyck2015}.
    
    \subsection{GPRI-based WDD processing of a single count}
    \vspace{3pt}
        \label{subsec:countwiseWDD1}
        
        Coming back to the arguments of section \ref{sec:theorysection1}, it now remains to determine what the WDD method can extract from individual electrons. For this purpose, a Dirac delta-function can be used as a basic building block to model the collapse of an electron wavefront at a specific location of the detector. Specifically, a single count experimental intensity, containing an isolated detection event at an arbitrary coordinate $\left[{\vec{r}_s}^{\,\,\,c}\,;\,{\vec{q}_d}^{\,\,\,c}\right]$, is given by
        \begin{equation}
            \label{eq:singlecountdata}
            I^{exp}_{\vec{r}_s}\left(\vec{q_d}\right) \, = \, \delta\left(\vec{r}_s-{\vec{r}_s}^{\,\,\,c}\right) \, \delta\left(\vec{q}_d-{\vec{q}_d}^{\,\,\,c}\right) \quad.
        \end{equation}
        Following arguments of subsection \ref{subsec:doseandsparsity}, this product of delta-functions should thus be taken as an imperfect estimate for the underlying $I^{det}_{\vec{r}_s}\left(\vec{q_d}\right)$.
        
        To facilitate further derivations, equation \ref{eq:singlecountdata} ignores the finite size of the detector pixels, which leads to a limited accuracy in selectable values of $\vec{q_d}$. In practice however, this ambiguity is solved by an explicit inclusion of $M\left(\vec{R}\right)$ in formulas \ref{eq:modulationtransferfunction} and \ref{eq:Wignerdistributions1}. Specifically, it is assigned a value of zero at reciprocal real-space coordinates $\vec{R}$ beyond a certain opening, which represents this smallest resolvable scattering vector interval. Pixel size can thus be accounted for through a restriction of used coordinates in a final summation step, as clarified below and equivalently to the process described in equation \ref{eq:directWDD}. Of course, this aspect only needs to be considered when the native sampling of detector space is less fine than implied by the maximum value of $\parallel\vec{R}\parallel$ in the employed numerical grid. Aside from those considerations, a perfect MTF \cite{Ruskin2013} is typically assumed.
        
        Based on equation \ref{eq:phasespacereformulation}, the phase space-reformulated scattering data is then equal to
        \begin{equation}
            \label{eq:singlecountphasespacedata}
            J_{\vec{Q}}\left(\vec{R}\right) \, \equiv \, e^{-i 2\pi \vec{Q}\cdot{\vec{r}_s}^{\,\,\,c}} \, e^{i 2\pi {\vec{q}_d}^{\,\,\,c}\cdot\vec{R}} \quad,
        \end{equation}
        which is simply a product of two plane wave components, i.e. translating their presence in the scan position-wise exit wave $\Psi_{\vec{r}_s}\left(\vec{r}_0\right)$ consistently with arguments given in subsection \ref{subsec:infofromsinglecounts}. Inserting formula \ref{eq:singlecountphasespacedata} within equation \ref{eq:directWDD} leads to an intermediary real-space result given by
        \begin{equation}
            \label{eq:singlecountWDDGPRI}
            \gamma \, T^{WDD}\left(\vec{r}\right) \, = \, G^{WDD}_{{\vec{q}_d}^{\,\,\,c}}\left(\vec{r}-{\vec{r}_s}^{\,\,\,c}\right) \quad,
        \end{equation}
        thus introducing a new library of guide functions and validating the hypothesis made in subsection \ref{subsec:GPRIproposal}. The notation $\gamma\,=\,\tilde{T}^{WDD^{\,*}}\left(\vec{0}\right)$ is introduced for simplification. In particular, this term acts as a proportionality constant and indicates the need for a normalization by the square rooted DC component, after processing the full scattering data, like in the conventional approach. This, however, does not preclude real-time observations and has no incidence on measurements of the phase shift.
        
        Most importantly, the guide functions are straightforwardly derived, and are defined in $\vec{Q}$-space by
        \begin{equation}
            \label{eq:WDDguidefunction}
            \tilde{G}^{WDD}_{{\vec{q}_d}}\left(\vec{Q}\right) \, = \, \sum\limits_{\vec{R}} \, \frac{ M\left(\vec{R}\right) \Gamma\left(-\vec{Q}\,;\,\vec{R}\right) \, e^{-i 2\pi \vec{q}_d\cdot\vec{R}} }{ \epsilon \, + \, \mid M\left(\vec{R}\right) \Gamma\left(-\vec{Q}\,;\,\vec{R}\right) \mid^2 }
        \end{equation}
        As such, the quantity $G^{WDD}_{\vec{q}_d}\left(\vec{r}\right)$ is shown to be complex and unitless. It results from an inversion of the known Wigner distribution $\Gamma\left(\vec{Q}\,;\,\vec{R}\right)$, with no information having been introduced about the specimen. Moreover, it encompasses a plane wave term $e^{-i 2\pi \vec{q}_d\cdot\vec{R}}$, outlining the available knowledge on activated scattering vectors. The full library can thus be calculated straightforwardly from pre-established illumination parameters, including the wavelength $\lambda$, convergence half-angle $\alpha$ and aberration function $\chi\left(\vec{q}_0\right)$. Finally, as explained above, the summation over $\vec{R}$ only goes as far as the opening of $M\left(\vec{R}\right)$ allows it, thus accounting for detector pixel size.
    
    \subsection{GPRI-based WDD processing of multiple counts}
    \vspace{3pt}
        \label{subsec:countwiseWDD2}
        
        From the basic concept formalized by equation \ref{eq:singlecountWDDGPRI}, it is straightforward to pursue with a reconstruction done with multiple detected electrons. In particular, starting from an event-based representation, the whole experimental dataset can be modeled as a vector of coordinates $\left[{\vec{r}_s}^{\,\,\,j}\,;\,{\vec{q}_d}^{\,\,\,j\,;\,k}\right]$ \cite{Jannis2022,Annys2025}. That is, for one arbitrary scan position $j\in\left[1\,;\,N_{s\,;\,x}\,N_{s\,;\,y}\right]$, multiple electrons of indices $k\in\left[1\,;\,n\left({\vec{r}_s}^{\,\,\,j}\right)\right]$ are received. Extending from formulas \ref{eq:singlecountdata} and \ref{eq:singlecountphasespacedata} then leads to
        \begin{equation}
        \begin{split}
            \label{eq:multicountdataandphasespacedata}
            I^{exp}_{\vec{r}_s}\left(\vec{q_d}\right) \, & \propto \, \sum\limits_{{\vec{r}_s}^{\,\,\,j}} \, \sum\limits_{{\vec{q}_d}^{\,\,\,j\,;\,k}} \, \delta\left(\vec{r}_s-{\vec{r}_s}^{\,\,\,j}\right) \, \delta\left(\vec{q}_d-{\vec{q}_d}^{\,\,\,j\,;\,k}\right) \\
            J_{\vec{Q}}\left(\vec{R}\right) \, & \equiv \, \sum\limits_{{\vec{r}_s}^{\,\,\,j}} \, \sum\limits_{{\vec{q}_d}^{\,\,\,j\,;\,k}} \, e^{-i 2\pi \vec{Q}\cdot{\vec{r}_s}^{\,\,\,j}} \, e^{i 2\pi {\vec{q}_d}^{\,\,\,j\,;\,k}\cdot\vec{R}} \quad.
        \end{split}
        \end{equation}
        As is noteworthy here, the experimental intensity is considered proportional to this superposition of Dirac delta-functions rather than simply equal to it, since an additional normalization step is needed as well.
        
        Based on equation \ref{eq:multicountdataandphasespacedata}, the calculation of a WDD result is then done by
        \begin{equation}
            \label{eq:multicountWDDGPRI1}
            \gamma \, T^{WDD}\left(\vec{r}\right) \, = \, \sum\limits_{{\vec{r}_s}^{\,\,\,j}} \, \frac{1}{n\left({\vec{r}_s}^{\,\,\,j}\right)} \, \sum\limits_{{\vec{q}_d}^{\,\,\,j\,;\,k}} \, G^{WDD}_{{\vec{q}_d}^{\,\,\,j\,;\,k}}\left(\vec{r}-{\vec{r}_s}^{\,\,\,j}\right) \quad,
        \end{equation}
        thus by a simple summation of individual $\vec{q}_d$-selected and $\vec{r}_s$-shifted guide functions $G^{WDD}_{\vec{q}_d}\left(\vec{r}\right)$, as was also devised in subsection \ref{subsec:GPRIproposal}. The division by $n\left({\vec{r}_s}^{\,\,\,j}\right)$ constitutes a pattern-wise normalization strategy, as was initially proposed in ref. \cite{Seki2018} and used in ref. \cite{LalandecRobert2025}. Its alternative would be a global approach, following
        \begin{equation}
            \label{eq:multicountWDDGPRI2}
            \gamma \, T^{WDD}\left(\vec{r}\right) \, = \, \frac{1}{N_{e^-}} \, \sum\limits_{{\vec{r}_s}^{\,\,\,j}} \sum\limits_{{\vec{q}_d}^{\,\,\,j\,;\,k}} \, G^{WDD}_{{\vec{q}_d}^{\,\,\,j\,;\,k}}\left(\vec{r}-{\vec{r}_s}^{\,\,\,j}\right) \quad,
        \end{equation}
        where the expectancy $N_{e^-}$ is first estimated by averaging the known values of $n\left(\vec{r}_s\right)$.
        
        Finally, as an alternative to a strict detection event-wise approach, GPRI can be extended back to the original continuous description of the acquired intensity through
        \begin{equation}
            \label{eq:continuousWDDGPRI}
            \gamma \, T^{WDD}\left(\vec{r}\right) \, \propto \, \sum\limits_{\vec{r}_s} \sum\limits_{\vec{q}_d} \,  I^{exp}_{\vec{r}_s}\left(\vec{q_d}\right) \, G^{WDD}_{\vec{q}_d}\left(\vec{r}-\vec{r}_s\right) \quad,
        \end{equation}
        with arbitrary normalization. This then shows the universality of the methodology with regards to data format. It is also interesting to note that this expression entails a convolutive formulation for the WDD process. From the mathematical viewpoint, this is consistent with the Fourier convolution theorem. As was noted in the introduction, another parallel can be made between equation \ref{eq:continuousWDDGPRI} and the OBF-STEM method \cite{Ooe2021,Ooe2023,Ooe2024,Ooe2026}, at least in the case where its is applied on single pixels of a DED. This is discussed in more details in subsection \ref{subsec:comparisontoOBF}.

\section{Implementation, interests and first application of the new methodology}
    
    \subsection{Practical workflow of the GPRI procedure}
    \vspace{3pt}
        
        Under the GPRI framework, analytical ptychography consists in a two-steps workflow. First, a library $G^{WDD}_{{\vec{q}_d}}\left(\vec{r}\right)$ is calculated. This is done through a process combining FFT and DFT, leading to a compact four-dimensional distribution which can be handled straightforwardly by a commercial GPU, at least in standard illumination conditions. Numerical details of this pre-calculation step are provided in appendix 1. Second, a reconstruction of \\$\gamma \, T^{WDD}\left(\vec{r}\right)$ is performed through a cumulative addition. This is ensured by a combined selection and shifting process of the priorly generated guide functions, directed by the 4D coordinates making up the scattering data, which is scanned through in a sequential approach.
        
        In treating an event-based dataset, the procedure relies on equations \ref{eq:multicountWDDGPRI1} or \ref{eq:multicountWDDGPRI2}, depending on the chosen normalization strategy. Hence, the contribution of each count is considered separately, though with some possible partitioning strategies. In the yet more conventional case of a frame-based dataset, this is equation \ref{eq:continuousWDDGPRI} that is employed. There, a separate GPU-accelerated Einstein summation is performed for each CBED pattern, along detector pixels $\vec{q_d}$ and relying on a pre-implemented tool \cite{Paszke2019}. Such usage of the einsum algorithm is mathematically equivalent to the series of single count-level $\vec{q_d}$-wise selections, as explained above. Thus, the dense representation involves measuring the complete $\vec{r}_s$-specific contributions, isolated from one another. Crucially, the reconstruction process is the same for both the post- and the in-acquisition cases. The workflow also reflects a prior innovation reported in ref. \cite{Annys2025}, in that the common treatment pipelines for MR-STEM data can be adapted to an event-represented dataset without the need to reform a collection of 2D CBED patterns in-memory.
        
        While this work focuses on the implementation of WDD ptychography by GPRI, the demonstration performed in subsections \ref{subsec:countwiseWDD1} and \ref{subsec:countwiseWDD2} can be replicated for SBI and iCoM. As such, formulations of guide functions exist for those two STEM-based phase retrieval methods as well. They are detailed in appendix 2 of this publication. Moreover, for SBI imaging, the same general advantages in terms of numerical complexity are found.
    
    \subsection{Numerical aspects and computational cost}
    \vspace{3pt}
        \label{subsec:numericalaspects}
        
        \begin{figure}
            \centering
            \includegraphics[width=0.8\columnwidth]{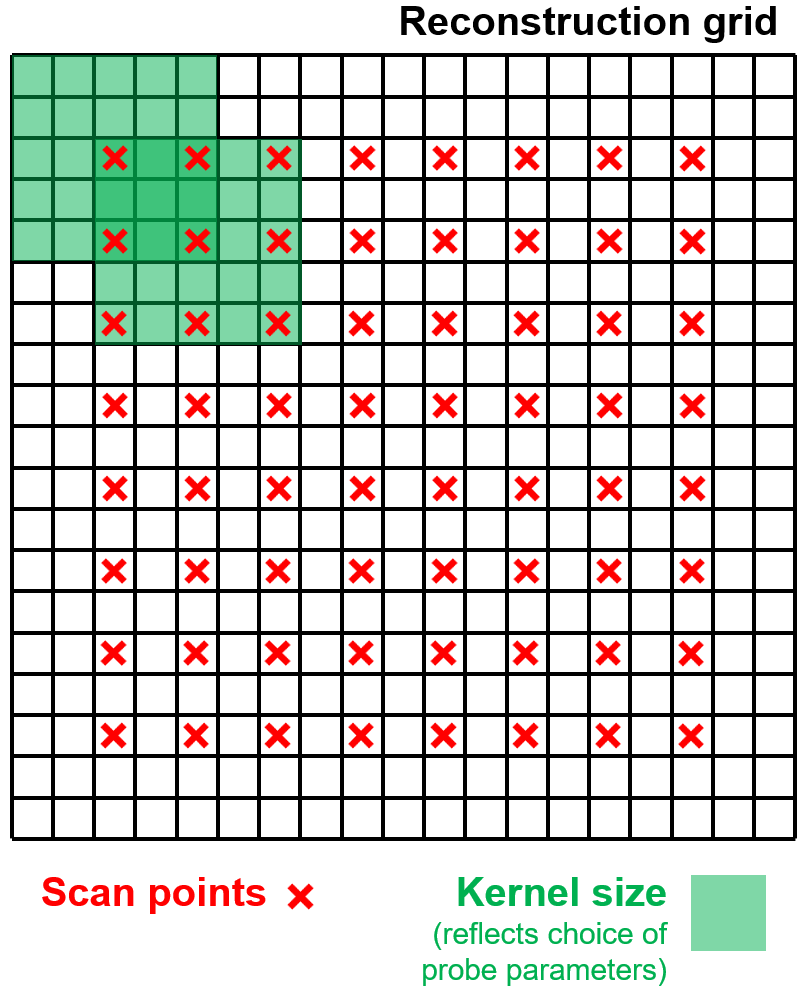}
            \caption{Illustration of the practical processing in GPRI. The kernel-limited guide function is used to update the few pixels of the reconstruction window that are found around the considered scan position $\vec{r}_s$. The reconstruction grid may be finer than the scan grid, and the ratio of the scan step over its pixel size is typically required to be an integer to permit straightforward selection of the updated pixels.}
            \label{fig:numericalsampling}
        \end{figure}
        
        As was briefly explained in subsection \ref{subsec:datapartitioning}, and formalized in \ref{subsec:phasespacedeconv}, the conventional WDD workflow \cite{Rodenburg1992,Li2014} involves Fourier transforming the data over the $\vec{r}_s$ dimensions to reach frequency space $\vec{Q}$, as well as a secondary $\vec{q}_d$-to-$\vec{R}$ transform. This is followed by a division by the illumination Wigner distribution, and by a $\vec{R}$-wise summation.
        
        When considering the dependence to image size, and more specifically to the amount of scan positions $N_s$, this algorithm entails a complexity \\$O\left( N_{s\,;\,x} \cdot N_{s\,;\,y} \cdot \log\left( N_{s\,;\,x} \cdot N_{s\,;\,y} \right) \right)$ for the typical FFT-based approach, and $O\left( {N_{s\,;\,x}}^2 \cdot {N_{s\,;\,y}}^2 \right)$ when using an DFT. To this is added the cost of the additional inverse FFT (iFFT) along detector space, which follows \\$O\left( N_{d\,;\,x} \cdot N_{d\,;\,y} \cdot \log\left( N_{d\,;\,x} \cdot N_{d\,;\,y} \right) \right)$ for a detector made of $N_d \, = \, N_{d\,;\,x} \, N_{d\,;\,y}$ pixels. Finally, the multiplication step along the four phase space dimensions may also be memory-intensive, as the inverted illumination Wigner distribution is expected to be already available for the procedure. The same is true for the complete reformulated dataset $J_{\vec{Q}}\left(\vec{R}\right)$.
        
        On the other hand, in a GPRI-based processing of the dataset, and still assuming a dense data format, the scan grid dependence of the total numerical complexity follows $O\left( N_{s\,;\,x} \cdot M_x \cdot N_{s\,;\,y} \cdot M_y \right)$. There, $M_x$ and $M_y$ are the pixel dimensions of the employed truncated kernel area, which has to be chosen large enough to contain the full $\vec{r}$-wise extension of $G^{WDD}_{{\vec{q}_d}}\left(\vec{r}\right)$. That is, in a single update of the reconstruction result around a probe location $\vec{r}_s$, only a total $M_x \, M_y$ of pixels are modified. This is illustrated in fig. \ref{fig:numericalsampling}. In practice, the updated areas overlap among distinct scan point-specific acquisitions, thus reflecting the working condition for ptychography in general \cite{Bunk2008,Edo2013,Huang2014}.
        
        The size of this for-processing kernel is fixed by the illumination conditions and does not depend on other parameters. As a consequence, enlarging the scanned area, e.g. by introducing a number $\Delta N_s$ of new points over the full raster grid, entails a equal increment in the total amount of calculations needed for the complete processing.
        
        In general, for the treatment of 2D frames, the needed collection of $\vec{r}_s$-wise Einstein summations plays the same role as a phase space-wise deconvolution. The new solution however offers a strong numerical advantage, at least over a DFT-based implementation, as \\$M_x \, M_y \, << \, N_{s\,;\,x} \, N_{s\,;\,y}$ in typical conditions. When comparing to the FFT-based case, this is less striking, since the situation where $M_x \, M_y \, < \, \log\left( \, N_{s\,;\,x} \, N_{s\,;\,y} \, \right)$ remains unrealistic, except in the prospect of achieving fields of view of several hundreds of nm. Nevertheless, other aspects of GPRI still make it an interesting alternative there, as is detailed in the rest of this subsection.
        
        The dimensional reduction mentioned above reflects a deeper physical aspect of how $\Gamma\left(\vec{Q}\,;\,\vec{R}\right)$ is represented in the new approach. That is, only a probe-localized segment of phase space needs to be considered to generate $G^{WDD}_{{\vec{q}_d}}\left(\vec{r}\right)$, as is explained in more details in appendix 1. This has a major implication with regards to memory requirements, in that the library of guide functions has a fixed, compact, size of $N_{d\,;\,x}\,N_{d\,;\,y}\,M_x\,M_y$ pixels in total, i.e. with no adaptation to the scanned area and a lower memory requirement. The diffraction data, whether it is made of frames or events, also does not need to be fully loaded at all time, and is not required to be pre-processed. For those reasons, GPRI can be very memory-efficient, in comparison to the alternatives.
        
        Furthermore, it should be noted the conventionally needed $\vec{q}_d$-to-$\vec{R}$ iFFT is not present at all in the new processing workflow. This is because it is equivalently represented as a distinct step within library calculation, though this is across a smaller 4D numerical grid. And, concerning the multiplication of $J_{\vec{Q}}\left(\vec{R}\right)$ by a priorly inverted $\Gamma\left(\vec{Q}\,;\,\vec{R}\right)$, its counterpart is directly found in the successive applications of the einsum algorithm, as was mentioned above.
        
        Crucially, this latter computational cost is where an additional, highly important, economy can be made. That is, by treating the scattered electrons as a collection of isolated incidences, rather than exhaustively extracting and summing the weighted contributions of each detector pixels, thus making full use of the count-wise formulation of GPRI.
        
        In the case of an event-based dataset, practical calculation steps consist in simpler sets of $M_x \, M_y$ additions, i.e. one per electron. There is thus no requirement for an intermediary $I^{exp}_{\vec{r}_s}\left(\vec{q_d}\right)$-weighted cumulation, and the total number of pixel updates equals the product between $M_x\,M_y$ and the amount of detection events obtained in the whole scan. The latter is also modulated by multiple counting effects \cite{Mir2017,Kuttruff2024}, though this may be corrected by a declustering algorithm \cite{VanSchayck2020,Jannis2022,Kuttruff2024}, inline or offline.
        
        As such, in the event-driven procedure, the individual, single count-level, operations are reduced to their simplest form, and the final numerical complexity scales with the beam current and the probe position dwell time directly, rather than with the size of the scan grid, at least in the strict sense. This also reflects the higher flexibility of an EDD with regards to acquisition time, e.g. allowing an arbitrary density in the scan grid, given a predetermined total acquisition time and dose \cite{Jannis2022}.
        
        As was explored in details in ref. \cite{Annys2025}, and reflecting prior work on sparsification approaches \cite{Pelz2022}, this solution can lead to a significantly higher processing speed than the treatment of equivalent 2D frames. More generally, the advantage is retained in a wide range of routine experimental conditions, and is especially striking in the low-dose case, which is also where analytical ptychography is most relevant \cite{Pennycook2019,OLeary2021,LalandecRobert2025}. Finally, the combination with a specialized event-driven processing pipeline, as was introduced in ref. \cite{Annys2025}, is another important aspect of the new workflow.
    
    \subsection{Real-space profiles of count-wise contributions}
    \vspace{3pt}
        \label{subsec:extractableinformation}
        
        \begin{figure*}
            \centering
            \includegraphics[width=1.0\textwidth]{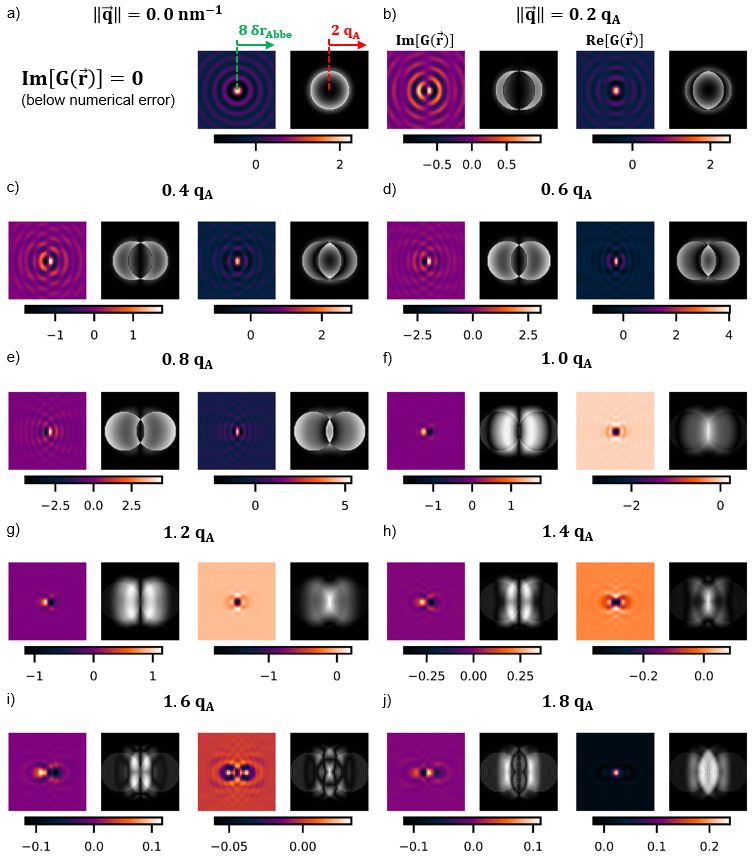}
            \caption{Depiction of a library $G^{WDD}_{\vec{q}}\left(\vec{r}\right)$ along $\vec{r}$, for a few indicated values of $\vec{q}\,=\,\left[\,q_x\,;\,0\,\right]$, equal to fractions of the aperture radius $q_A$. No MTF is included in the calculation, hence the use of the scattering vector dimension directly. For each case, both the imaginary and real parts of the guide function are shown, alongside their Fourier transform amplitude. An exception is the case of a), i.e. for $\parallel\vec{q}\parallel\,=\,0\,\text{nm}^{-1}$, where the imaginary part is found below numerical precision. The real-space extent shown covers a radius equal to $8\,\delta r_{Abbe}$ and frequency space is diffraction-limited. Colorbars reflect numerical, and unitless, values taken by the guide functions. The Wiener filter parameter was $\epsilon\,=\,10^{-6}$.}
            \label{fig:guidefunctiondepiction}
        \end{figure*}
        
        From the theoretical standpoint, one interest of the GPRI framework is the possibility it offers to make predictions on future measurement responses. Specifically, since each element making up a reconstruction is taken from $G^{WDD}_{\vec{q}_d}\left(\vec{r}\right)$, the contribution of a specific detector coordinate $\vec{q}_d$, both along $\vec{r}$- and $\vec{Q}$-space, is known. As a first demonstration of this principle, selected guide functions are displayed in fig. \ref{fig:guidefunctiondepiction}, for arbitrary illumination parameters and under a real-space radius of $8\,\delta r_{Abbe}$. The chosen Wiener parameter was $\epsilon\,=\,10^{-6}$, as used in ref. \cite{LalandecRobert2025}, and the employed scattering vectors $\vec{q}\,=\,\left[\,q_x\,;\,0\,\right]$ were given linearly increasing moduli with an horizontal vectorial direction. The notation is justified by the absence of an MTF in the calculation.
        
        Generally, fig. \ref{fig:guidefunctiondepiction} shows that each count-wise contribution possesses a specific, and detailed, real-space profile, leading to a non-obvious influence on the final measured phase shift map $\varphi^{WDD}\left(\vec{r}\right)$. As is reflective of an Airy disk-like probe amplitude, those guide functions also display a radial dampening accompanied by an oscillatory behavior. Interestingly, those medium-to-long-range oscillations present a $\vec{q}$-dependent frequency, giving rise to negative and positive values in both the imaginary and real parts. This implies that a GPRI processing, and its associated noise formation mechanism, does not only consist in a straightforward cumulation of image features, but rather gives rise to a fine interplay of additive and subtractive effects along a progressive treatment of scattering data.
        
        Whereas the real component of a contribution is point-symmetric, its imaginary part is point-antisymmetric. As such, $G^{WDD}_{\vec{q}_d}\left(\vec{r}\right)$ expectedly shows trigonometric characteristics. In correlation with the argument made on the radial oscillations, this means that two counts received at coordinates with identical moduli and opposite vectorial directions would cancel each other's effect on \\$Im\left[\,\gamma\,T^{WDD}\left(\vec{r}\right)\,\right]$, while carrying identical information on $Re\left[\,\gamma\,T^{WDD}\left(\vec{r}\right)\,\right]$. Accordingly, an acquisition done over vacuum, providing diffraction patterns equal to the aperture profile $A^2\left(\vec{q}\right)$, would consistently lead to a real-valued measurement of the transmission function, i.e. showing no retrievable phase shift.
        
        This also implies that, in order to accurately access the underlying $\varphi\left(\vec{r}_0\right)$, a sufficient azimuthal sampling is required among accessible scattering vectors, so that the dominant directionality of the local momentum redistribution can be elucidated \cite{Muller2014}. A similar notion exists in the case of differential phase contrast methods, as discussed e.g. in ref. \cite{Muller-Caspary2018a,Pollath2021,Li2022,Grieb2024}.
        
        Moreover, the $\vec{q}$-coordinates which provide the highest range of variations in $\gamma\,T^{WDD}\left(\vec{r}\right)$, and which are thus expected to offer the most noise-robust information on the specimen, are those with a modulus closest to the primary beam radius. The concerned guide functions also possess the most centrally localized profiles. Relatedly, the scattering angles fulfilling $\parallel\vec{q}\parallel\,=\,q_A$ correspond to the highest gradient in the intensity distribution $I_{\vec{r}_s}\left(\vec{q}\right)$, and thus constitute the points where interaction-induced momentum redistributions are the most visible.
        
        This latter finding is reminiscent of the work reported in ref. \cite{Lorenzen2024}, where Fourier ptychography-like \cite{Zheng2021} reconstructions were demonstrated while solely making use of tilt angles close to the numerical aperture, i.e. limiting acquisitions to the most influential propagation directions only. Here, by argument of optical reciprocity \cite{Cowley1969a,Krause2017}, those would correspond to near-$q_A$ scattering vectors. Hence, the guide function profiles depicted in fig. \ref{fig:guidefunctiondepiction} tend to validate the theoretical foundation of this tilted-wave methodology, in that a good dose-efficiency can be expected, and show relevance for current developments in the field \cite{Zhao2025}. This nevertheless also indicate a possible limitation. That is, $\vec{q}$ values above or well-below this limit provide rich information as well, contributing to a more complete and accurate recovery of the full frequency spectrum of the specimen.
        
        Accordingly, it can be observed that all the contributions have a different profile in $\vec{Q}$-space as well, with a point-symmetric amplitude. While the near-$q_A$ cases inform on the highest retrievable image frequencies, reflecting their spatial localization, the low-$\vec{q}$ ones give rise to long-range effects in the reconstruction, though without a restriction to small $\vec{Q}$ values specifically. This is because their dominant oscillative components are themselves resolved with a certain spatial frequency, as is exemplified e.g. by the ring-like dependence in the $\parallel\vec{q}\parallel\,=\,0\,\text{nm}^{-1}$ case. As a result, an excessively reduced selection of used scattering vectors, in the accessible $I^{det}_{\vec{r}_s}\left(\vec{q}_d\right)$, can be expected to lead to a modification of frequency transfer. This was implicitly demonstrated by recent work making use of special detectors geometries \cite{Kim2023,Lei2025}.
    
    \subsection{Reconstruction test on simulated sparse data}
    \vspace{3pt}
        \label{subsec:testonsim}
        
        \begin{figure}
            \centering
            \includegraphics[width=0.9\columnwidth]{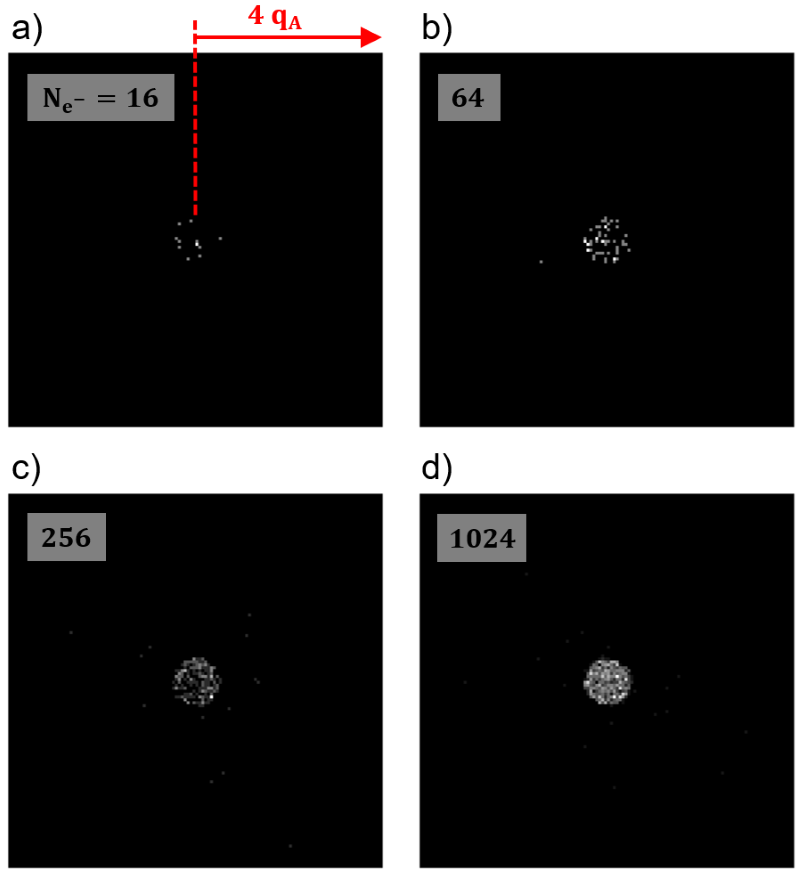}
            \caption{Simulated CBED patterns, after a 4-by-4 binning step leaving a frame of 128$^{\text{2}}$ pixels. The range of scattering vectors covered goes up to 4 times the primary beam radius $q_A$, as indicated in a). A dose-limitation strategy was employed, following Poisson statistics and producing sparse data. This was ensured under expectation values $N_{e^-}$ of a) 16, b) 64, c) 256 and d) 1024.}
            \label{fig:simpatterns}
        \end{figure}
        
        \begin{figure*}
            \centering
            \includegraphics[width=1.0\textwidth]{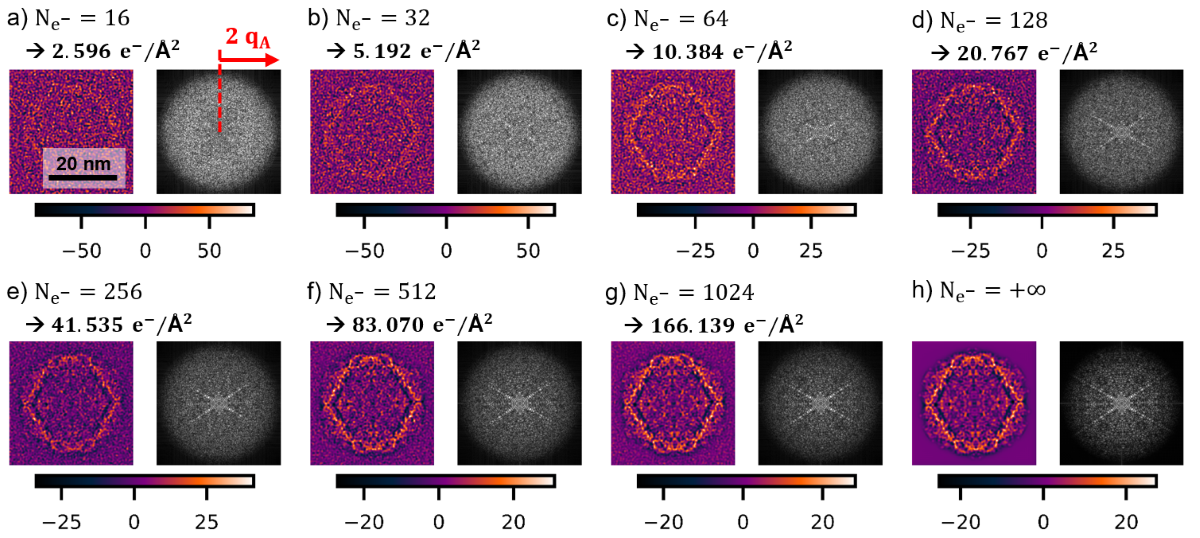}
            \caption{Results of GPRI-based WDD imaging, employing simulated CBED patterns. The specimen is a poliovirus, illuminated under an acceleration voltage of 200 kV and a convergence half-angle $\alpha\,=\,2\,\text{mrad}$. Calculations are done for a variety of average numbers of electrons per pattern $N_{e^-}$, and corresponding doses given in $e^-/\text{\AA}^2$. For each case, the position-dependent measurement of the projected potential, denoted $\mu^{WDD}\left(\vec{r}\right)$, is displayed alongside the square root of its Fourier transform's amplitude, i.e. $\sqrt{\mid\tilde{\mu}^{WDD}\left(\vec{Q}\right)\mid}$. The colorbars are in V$\cdot$nm.}
            \label{fig:demonstration}
        \end{figure*}
        
        In order to validate the applicability of the method for a relevant test case, a simulation of CBED patterns was conducted on a poliovirus specimen, using the AbTEM software \cite{Madsen2021a} and a structure that was originally recovered in the work reported in ref. \cite{Basavappa1994}. Specifically, the propagation of an electron wave through its volume was modeled using the multislice algorithm \cite{Cowley1957,Goodman1974,Ishizuka1977}, while employing an acceleration voltage of 200 kV and a convergence half-angle $\alpha\,=\,2\,\text{mrad}$.
        
        A slice thickness of 400 pm was chosen, which is sufficient to accurately represent the dynamical diffraction behavior under those conditions \cite{Leidl2023,Leidl2025}. The parameterization of atomic potentials followed ref. \cite{Kirkland2020}. In parallel, the role of thermal diffuse scattering was ignored, as it is expected to have a negligible contribution in the concerned low angular range. For simplicity, the influence of an amorphous ice layer, usually embedding this type of biological specimens \cite{Dubochet1988}, was left out as well. Its importance in phase contrast methods was nevertheless explored e.g. in ref. \cite{Baxter2009,Leidl2023,LalandecRobert2025,Leidl2025}.
        
        Sampling in reciprocal space was ensured with a pixel size of about 0.0228 nm$^{\text{-1}}$, increased to 0.0913 nm$^{\text{-1}}$ pre-processing by a 4-by-4 binning step. Notably, following ref. \cite{Yang2015b,Susi2025} and the explanation provided in subsection \ref{subsec:countwiseWDD1} on the role of scattering vector accuracy, this binning step is not expected to cause issues in the ptychographic phase retrieval.
        
        The simulated experiment consisted in a scan of 144 by 144 positions with a lateral interval of 250 pm, permitting an overlap ratio of 78.7\% among the areas covered in neighboring acquisitions. No MTF was included. Dose-limitation was ensured post-simulation, on the binned frames, through the approach reported in ref. \cite{LalandecRobert2025}, i.e. under a random pixel selection repeated a number of times $n\left(\vec{r}_s\right)$, itself satisfying Poisson statistics. This then leads to sparse frames faithfully modeling quantization effects. Expectancies $N_{e^-}$ in the number of counts per pattern were given values of 16 to 1024, leading to a variety of doses. Examples of the resulting binned and dose-limited patterns are available in figure \ref{fig:simpatterns}, for a single scan position located at the edge of the virus, thus reflecting maximum scattering condition. The final case of $N_{e^-}\,=\,+\infty$ corresponded to a direct usage of the simulated $I_{\vec{r}_s}\left(\vec{q}\right)$.
        
        WDD retrieval of the projected potential $\mu^{WDD}\left(\vec{r}\right)$, expressed in V$\cdot$nm, was performed using the GPRI framework under a Wiener parameter $\epsilon\,=\,10^{-6}$, as used in subsection \ref{subsec:extractableinformation}. Results are provided in fig. \ref{fig:demonstration}.
        
        With regards to the prior generation of $G^{WDD}_{\vec{q}}\left(\vec{r}\right)$, following the process explained in appendix 1, the kernel radius of the initial calculation windows, i.e. both along $\vec{r}$ and $\vec{R}$, were given values of $16\,\delta r_{Abbe}$. The final $\vec{r}$-wise cutoff, limiting the updated region around scan positions $\vec{r}_s$, was done under $8\,\delta r_{Abbe}$. This latter step was accompanied by a multiplication with a Hann window, reaching zero at the radial edge of the conserved area. Furthermore, the simulated counts were treated under a scattering vector limit of $4\,q_A$, making use of an extensive dark field range. Under this somewhat demanding set of parameters, an optimal reconstruction accuracy could be ensured, avoiding spatial restriction-induced modifications of frequency transfer \cite{Yu2022a}. More generally, in practical applications, a variety of choices can be made to balance this aspect with processing speed and memory usage.
        
        Importantly, library calculation was performed a single time for all the dose-limited datasets. This is reflective of real experimental conditions, where permanently available libraries could be generated once for a collection of prospective measurement conditions and used as needed for real-time processing at the microscope. Concerning data treatment, the frame-based approach described by equation \ref{eq:continuousWDDGPRI} was employed in all cases, both for simplicity with regards to the native data format and to ensure a streamlined treatment.
        
        As expected for this analytical solution, the WDD results depicted in fig. \ref{fig:demonstration} show a remarkable dose-efficiency. In particular, interpretable image contrast, with a clear recovery of the internal features of the poliovirus, is obtained under a dose as low as 10 $e^-/\text{\AA}^2$. As such, it is instructive to compare those micrographs to e.g. the cryo-TEM experiments reported in ref. \cite{Butan2014}, where a comparable dose of 15 $e^-/\text{\AA}^2$ was used. In general, obtaining higher contrast with the same or a lower range of $N_{e^-}$ would also be possible through a reduction of numerical aperture, at a possible cost of spatial resolution \cite{Pei2023,Mao2024,Li2025,LalandecRobert2025}.
        
        Most importantly, those reconstruction results confirm the applicability of GPRI, and thus validate its underlying quantization-based description, here showing the same stability against noise as demonstrated by existing workflows \cite{OLeary2020,OLeary2021,LalandecRobert2025,Dearg2025}. As a side-note, the recovered phase shift map is found to possess a range of values covering about 0.4 rad, thus not fulfilling the weak phase object approximation, as was similarly argued for apoferritin in ref. \cite{LalandecRobert2025}. For additional reference, appendix 3 provides a side-by-side comparison of a GPRI-based WDD result with an SFPA-based reconstruction.

\section{Prospects and experimental demonstration of the GPRI framework}
    \label{sec:experiments}
    
    \subsection{Applicability for real-time imaging - Silicalite-1 zeolite}
    \vspace{3pt}
        \label{subsec:expdemozeolite}
        
        \begin{figure}
            \centering
            \includegraphics[width=0.9\columnwidth]{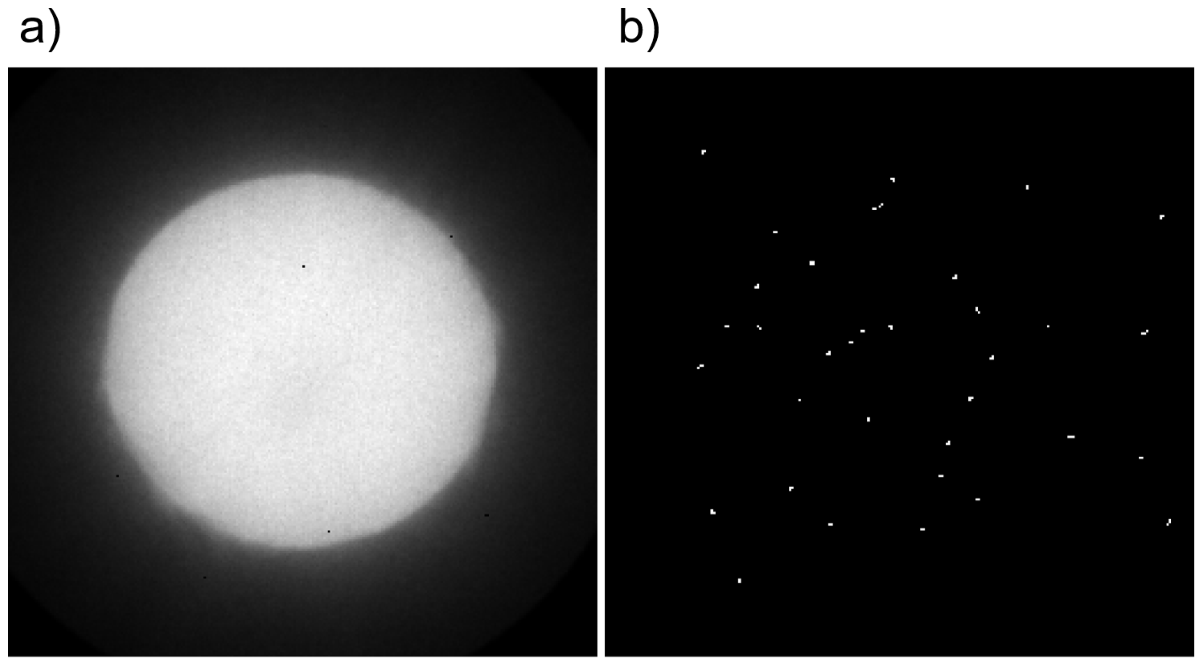}
            \caption{Reformation of 2D diffraction frames from the experimental event-based data, recovered from the zeolite specimen. a) PACBED pattern, b) single CBED pattern.}
            \label{fig:exppatternzeolite}
        \end{figure}
        
        \begin{figure*}
            \centering
            \includegraphics[width=1.0\textwidth]{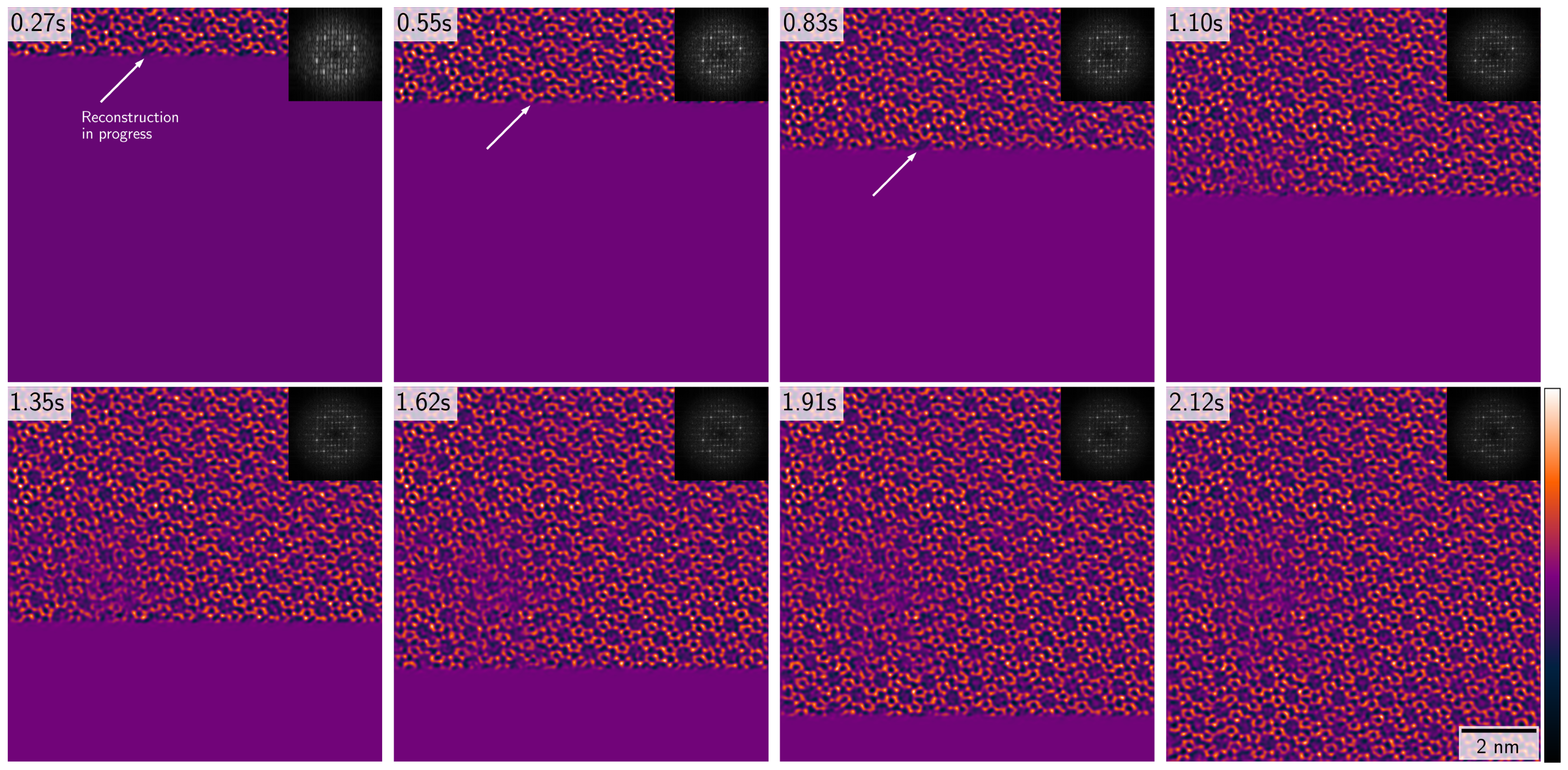}
            \caption{Result of an event-driven WDD reconstruction, done via the GPRI framework, applied on the silicalite-1 zeolite specimen. Illumination parameters include an acceleration voltage of 200 kV, a convergence semi-angle of 12 mrad, an approximate dose of 4200 $e^-/\text{\AA}^2$ and a scan step of about 10 pm. The figure depicts the quantity $Im\left[\,\gamma\,T^{WDD}\left(\vec{r}\right)\,\right]$, with a specific micrograph provided after each 1/8-portion of all scan positions has been treated. The required processing times are indicated as well, and the square rooted Fourier transform amplitude is given as an inset.}
            \label{fig:realtimeprocesszeolite}
        \end{figure*}
        
        With a first demonstration of the methodology having been done on simulated diffraction patterns in subsection \ref{subsec:testonsim}, thus showing good performance in recovering phase shift maps at low doses, it remains to investigate real-time imaging capacities, as well as the general applicability in routine work. To this end, more reconstructions were performed on experimental event-driven data, recovered using an aberration-corrected Titan Themis 60-300 instrument (Thermo Fisher Scientific). This microscope is equipped with an AdvaPIX TPX3 detector (Advacam), in an arrangement that was described in details in ref. \cite{Jannis2022}. This includes a custom retractable port, a supplementary scan engine and a reference clock with a phase-locked loop circuit. Some recent improvements, especially with regards to data handling, were also reported in ref. \cite{Annys2025}.
        
        A first processing test was conducted using patterns that were investigated prior, in ref. \cite{Annys2025}. The illuminated specimen was a silicalite-1 zeolite, i.e. a nanoporous material with applications in gas adsorption and catalysis \cite{Meenu2025}. As such, this first demonstration offers an appropriate reference for comparisons to existing work \cite{Jannis2022}, as well as a straightforward baseline for medium-resolution purposes, not yet requiring atomic accuracy.
        
        The dataset encompasses a square grid of 1024$^{\text{2}}$ scan positions, with a dwell time of 6 $\mu$s, covering a field of view of about 10 nm. The convergence half-angle was set to 12 mrad and the acceleration voltage to 200 kV. A beam current value of 1.1 pA led to a total dose of about 4200 $e^-/\text{\AA}^2$ in the measurement. Both those parameters were directly estimated from the data, while accounting for the priorly calibrated \cite{Jannis2022} degree of multiple counting \cite{Mir2017,VanSchayck2020}. No prior declustering \cite{VanSchayck2020,Jannis2022} step was performed, which is not expected to be problematic for the measurement, when considering the size of the focused probe and following the explanations provided in subsection \ref{subsec:countwiseWDD1} and appendix 1. To illustrate the area covered by the primary beam on the detector, a position-averaged CBED (PACBED) pattern, i.e. the sum of all frames reformed from detection events, is provided in figure \ref{fig:exppatternzeolite}. Alongside it, a single scan point-specific acquisition is shown as well.
        
        From the experimental events, a GPRI-based WDD reconstruction was conducted offline, with results provided in fig. \ref{fig:realtimeprocesszeolite}. This consists of snapshots corresponding to specific stages of the processing, i.e. after treating 1/8-portions of the scan array. For each intermediary micrograph, the total time invested in the calculation is provided. Notably, in this example, this is the imaginary part of the recovered transmission function that is shown, rather than an extracted phase shift. This is to reflect a prospective experimental situation, where the user is given direct access to $\gamma\,T^{WDD}\left(\vec{r}\right)$, being incremented at each new event, packet of events or scan position. Finally, the square root of the Fourier transform amplitude is given as an inset, like in fig. \ref{fig:demonstration}, to illustrate the recovery of specimen frequencies.
        
        For the initial generation of the guide functions, the radii of the calculation and for-processing kernels were respectively set to 8 and 4 times $\delta r_{Abbe}$, following the explanation given in appendix 1. The Wiener parameter was given a value $\epsilon\,=\,10^{-3}$. In the successive treatment of events, a 4-by-4 binning step of pixel coordinates was also done, thus reducing the effective detector dimensions $N_{d\,;\,x}\,/\,N_{d\,;\,y}$ from 256-by-256 to 64-by-64. As has been discussed in the simulation case, this standard practice \cite{Yang2015b,Susi2025} is not expected to create issues, and leads to an additional reduction of processing time. A Hann window was applied on the guide functions, like in subsection \ref{subsec:testonsim}. Finally, the size of the for-processing kernel, in pixels, was $\left[\,M_x\,;\,M_y\,\right] \, = \, \left[\,86\,;\,86\,\right]$. This has implications for the computational cost of the whole procedure, as explained in subsection \ref{subsec:numericalaspects}.
        
        Notably, while those kernel dimensions are relatively high, this can be explained by the requirement for an integer-valued ratio of the scan step over the reconstruction pixel size, as is explained in appendix 1. That is, if the scan sampling is finer than the diffraction limit of the measurement, it ends up directly defining the reconstruction window and may accordingly inflate the kernel dimensions, at least to some extent. This also corresponds to a condition of very high overlap of illuminated areas, which is here found equal to 98.7\% following the cross-correlation criterion used e.g. in ref. \cite{LalandecRobert2025}.
        
        On that last point, it should nevertheless be highlighted that, in an event-driven GPRI reconstruction, the scan grid is not the most relevant factor for the numerical complexity. As was explained in details in subsection \ref{subsec:numericalaspects}, the total number of operations needed is proportional to the amount of electrons in the dataset, without a direct role for the number of scan points. In other words, and considering equivalent illuminated surfaces and doses, whether the events are distributed among 1024$^{\text{2}}$ or, say, 256$^{\text{2}}$ positions should not make a difference in an event-driven ptychographic process. That is, assuming perfect stability of the specimen and without considering the sampling in conventional STEM images, produced in parallel to assist the operator.
        
        Implicitly, this also means that an oversampling of real-space, if needed for specific applications in the future, can be tolerated with no issues in this methodology, even with an added numerical cost due to requiring finer reconstruction window and guide functions. As a prospect, lifting this measurement limitation could, for instance, allow an improved control over the loss of spatial coherence related to vibrations and drift of the specimen stage, or to a small random jitter in the scan system, due to reducing the equivalent integration time of a CBED pattern \cite{VanDyck2011,Oxley2020}.
        
        Continuing, the explicit event-wise processing was ensured through the architecture described in ref. \cite{Annys2025}. The calculation was performed using a 16 core AMD Ryzen threadripper pro 5995wx CPU and an Nvidia RTX4090 GPU. Hence, this first test was performed while only relying on consumer devices, in contrast to other emerging work on the application of high-performance computing facilities in MR-STEM and ptychography \cite{Jones2022,Mukherjee2022,Wang2022c,Welborn2024}.
        
        For this dataset, the complete processing required a duration of 2.12 s, leading to a satisfactory high-resolution micrograph providing detailed information on the zeolite structure. Hence, the time taken, on average, by the treatment of a single scan position is found equal to 2.02 $\mu$s, i.e. a third of the dwell time. This clearly demonstrates that the GPRI algorithm, when embedded in an appropriate event processing architecture, is fully able to compete with the speed of the measurement setup. As the latter is representative of routine low-dose STEM work, this is very encouraging for a future implementation of live experimental feedback at the microscope \cite{Strauch2021,Yu2022a,Bangun2023,Weber2024}.
        
        Finally, it is interesting to note, again, the absence of a prior declustering of the event-based data \cite{VanSchayck2020,Kuttruff2024}, which would have further reduced the number of counts to be included in the calculation. This choice was made because, in effect, inserting this intermediary step in a future in-acquisition processing pipeline could reduce the speed of the complete reconstruction. Specifically, it is not fully clear whether eliminating the cost of processing multiply-counted electrons is worth this additional pre-processing of equivalent CBED patterns, when considering the final calculation time invested in each scan position. Hence, this perspective was not explored in this work.
    
    \subsection{Atomically resolved measurement - SrTiO$_{\text{3}}$ membrane}
    \vspace{3pt}
        \label{subsec:expdemoSTO}
        
        \begin{figure}
            \centering
            \includegraphics[width=0.9\columnwidth]{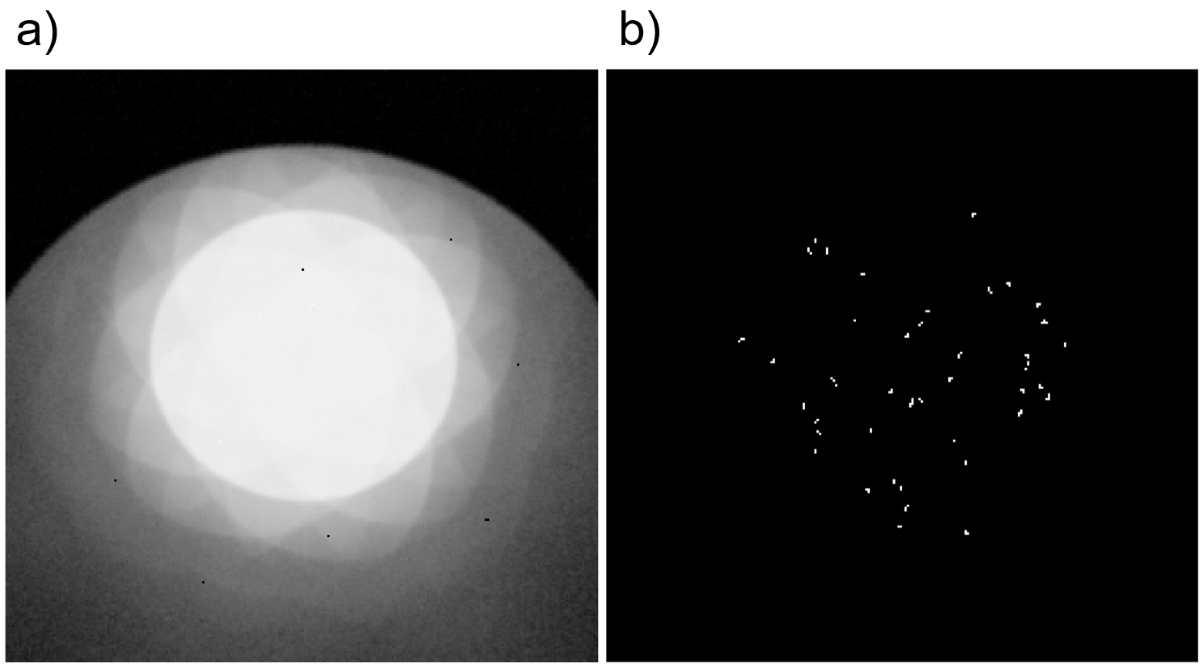}
            \caption{Reformation of 2D diffraction frames from the experimental event-based data, recovered from the SrTiO$_{\text{3}}$ specimen. a) PACBED pattern, b) single CBED pattern.}
            \label{fig:exppatternSTO}
        \end{figure}
        
        \begin{figure*}
            \centering
            \includegraphics[width=1.0\textwidth]{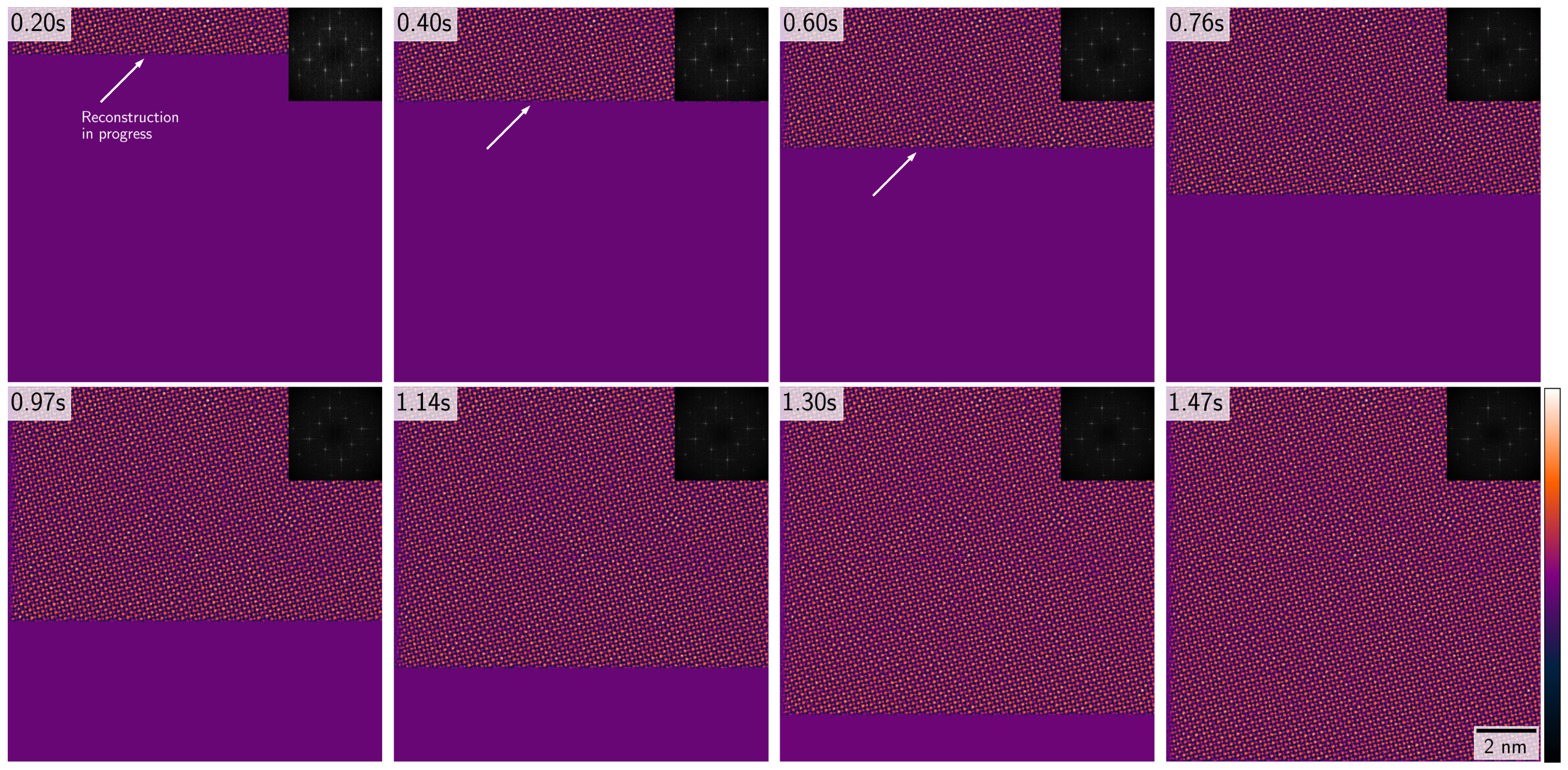}
            \caption{Result of an event-driven WDD reconstruction, done via the GPRI framework, applied on the SrTiO$_{\text{3}}$ specimen. Illumination parameters include an acceleration voltage of 200 kV, a convergence semi-angle of 21 mrad, an approximate dose of 2800 $e^-/\text{\AA}^2$ and a scan step of about 12 pm. The figure depicts the quantity $Im\left[\,\gamma\,T^{WDD}\left(\vec{r}\right)\,\right]$, with a specific micrograph provided after each 1/8-portion of all scan positions has been treated. The required processing times are indicated as well, and the square rooted Fourier transform amplitude is given as an inset.}
            \label{fig:realtimeprocessSTO}
        \end{figure*}
        
        As a second use case, a reconstruction was done using an epitaxial membrane of crystalline SrTiO$_{\text{3}}$. Its thickness is expected to be close to 10 nm, and it is oriented along the [100] zone axis. The acquisition was performed using the same instrument and experimental setup as presented in subsection \ref{subsec:expdemozeolite}, and involved 1024$^{\text{2}}$ scan positions, a dwell time of 5 $\mu$s, a field of view of 12.5 nm, an acceleration voltage of 200 kV and a convergence half-angle of 21 mrad. Given an estimated beam current of 1.3 pA, the final dose was 2800 $e^-/\text{\AA}^2$. Those measurement conditions lead, again, to a high value of area overlap, here reaching 94.0\%.
        
        Like in the last case, a PACBED pattern and a single reformed CBED frame are provided in fig. \ref{fig:exppatternSTO}. As it was more practical in this particular series of acquisitions, the high-angle annular dark field (HAADF) detector was left inserted, which leads to a visible shadow over the Timepix3 chip. This implies a minor underestimation in value, for the dose and beam current mentioned above. With regards to the reconstruction itself, this is not problematic, as the angular range included in the calculation can be reduced at will, here excluding the concerned area of reciprocal space. A few more details on this topic are provided in appendix 1.
        
        The results of an event-driven GPRI processing are provided in figure \ref{fig:realtimeprocessSTO}, following the same format as fig. \ref{fig:realtimeprocesszeolite}. That is, a progressively completed recovery of \\$Im\left[\,\gamma\,T^{WDD}\left(\vec{r}\right)\,\right]$, with indication of the time taken by partial processing. Choices of reconstruction parameters, approach for data handling and computing hardware were identical to those detailed in subsection \ref{subsec:expdemozeolite}.
        
        Importantly, the procedure was fully completed in 1.47 s. Such a reduction in processing time, compared to the zeolite case, is a consequence of the lower invested dose, and leads to an average duration of 1.40 $\mu$s for the treatment of a single scan position. This, again, is highly encouraging for a future real-time implementation at the microscope, and serves as proof that GPRI is appropriate for fast analytical ptychographic measurement, with atomic accuracy.
        
        Concerning this latter objective, it should also be pointed out that the micrographs presented in fig. \ref{fig:realtimeprocessSTO} offer high contrast and resolution, and would allow a direct characterization of projected crystalline features in the specimen, for instance. In particular, the spatial frequencies belonging to the SrTiO$_{\text{3}}$ lattice are clearly visible in the Fourier transform.
        
        This level of image quality would be appropriate for a wide variety of applications, especially given the relatively low electron dose invested. Moreover, what makes this particularly interesting, when considering the standard practices in the field, is the absence of a numerical compensation of residual aberrations in the optical system. That is, only a perfect, unaberrated, illumination was considered in the library calculation step.
        
        This highlights that a good, or at least sufficient, performance of the imaging method may be achievable without an offline correction step \cite{Thibault2009,Maiden2009,Yang2016,Li2025a,Varnavides2026}, i.e. only relying on the efficacy of modern corrector devices \cite{Batson2002,Erni2009,Morishita2018a}, and on their improving stability \cite{Barthel2013}. That much, of course, should not be surprising, given that STEM methods providing coherent contrast \cite{Okunishi2009,Findlay2009a,Muller2014,Muller-Caspary2018,Yucelen2018}, following the same elastic interactions as exploited in electron ptychography, have long shown that ability.
        
        Finally, it is interesting to note the presence of a subtle crystallographic deviation, over the full micrograph. Specifically, in the projected lattice unit, one of the column is shifted from the center. This is not caused by a mistilt of the crystal, as confirmed by fig. \ref{fig:exppatternSTO}.a, and could indicate some fine polarization effect in the material. This topic is left to future investigations.
    
    \subsection{Robustness to high thickness - NaCl nanoparticle}
    \vspace{3pt}
        \label{subsec:expdemoNaClparticle}
        
        \begin{figure}
            \centering
            \includegraphics[width=0.9\columnwidth]{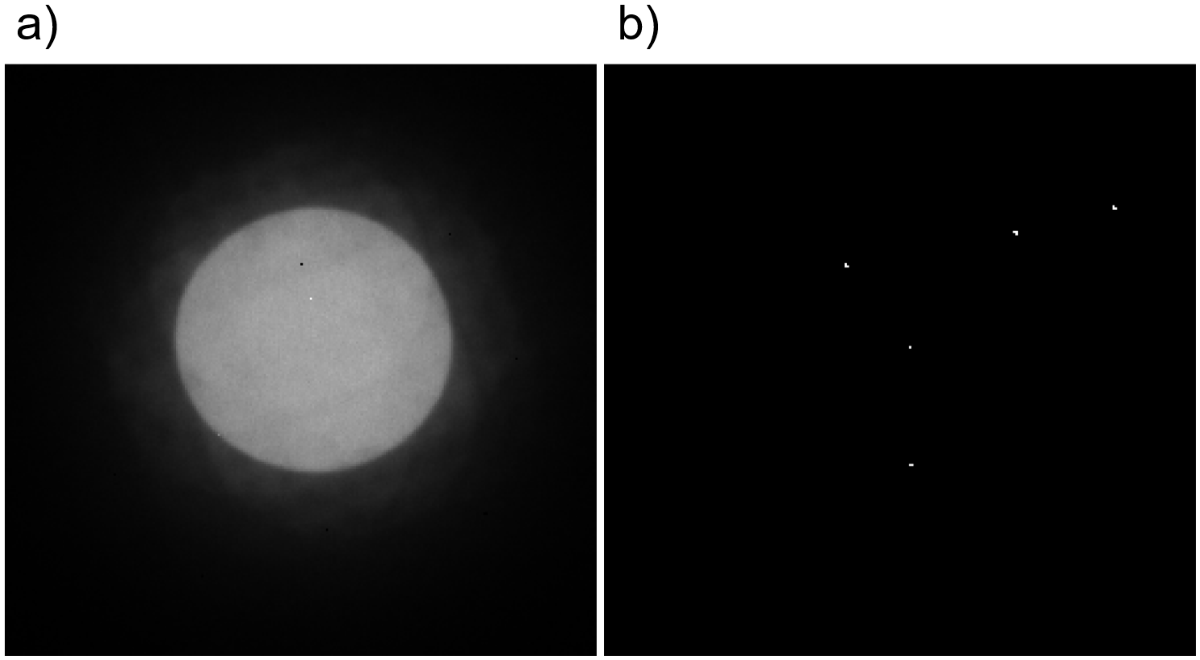}
            \caption{Reformation of 2D diffraction frames from the experimental event-based data, recovered from the NaCl nanoparticle specimen. a) PACBED pattern, b) single CBED pattern.}
            \label{fig:exppatternNaClparticle}
        \end{figure}
        
        \begin{figure*}
            \centering
            \includegraphics[width=1.0\textwidth]{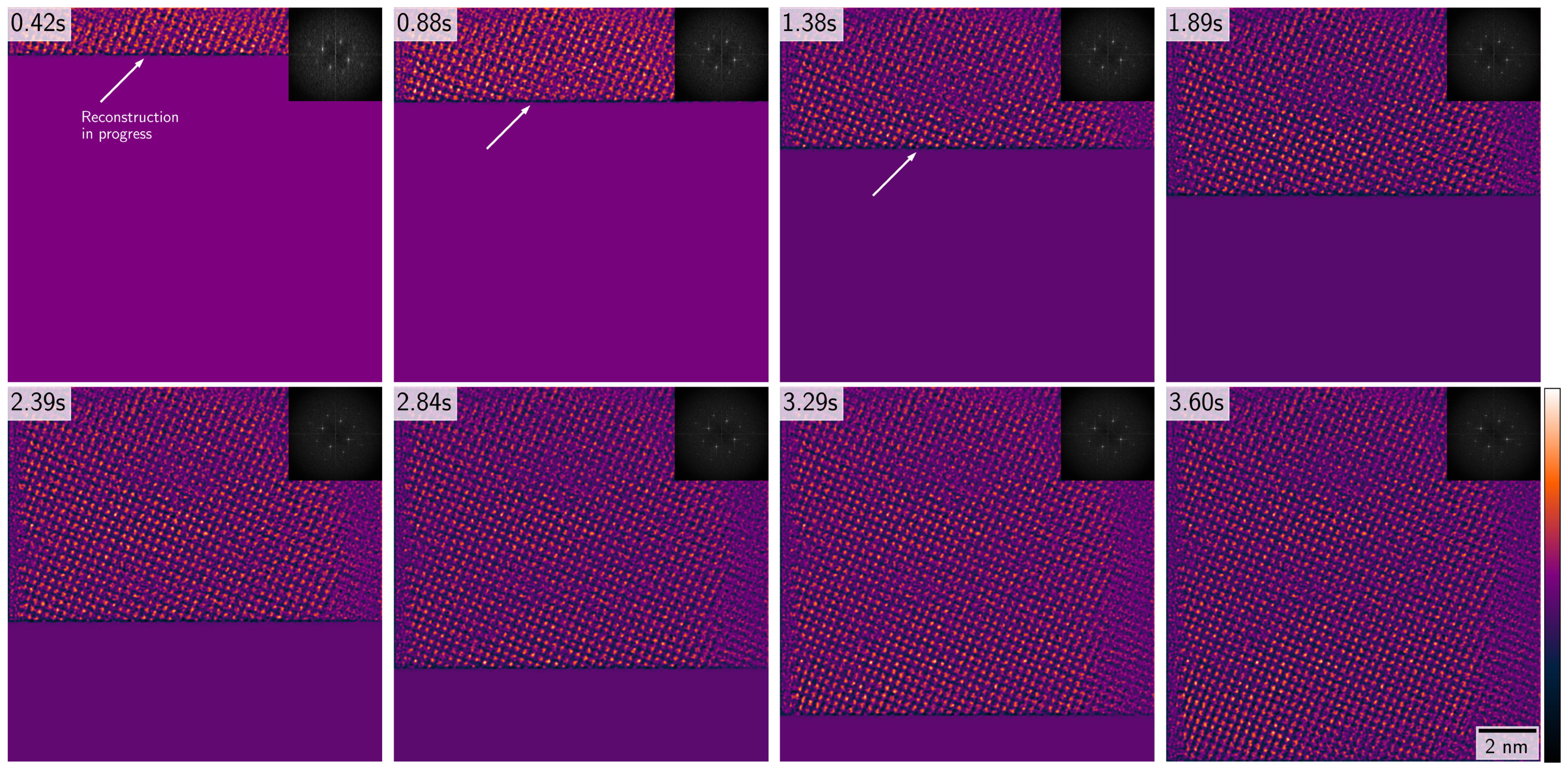}
            \caption{Result of an event-driven WDD reconstruction, done via the GPRI framework, applied on the NaCl nanoparticle specimen. Illumination parameters include an acceleration voltage of 200 kV, a convergence semi-angle of 18 mrad, an approximate dose of 1800 $e^-/\text{\AA}^2$ and a scan step of about 6 pm. The figure depicts the quantity $Im\left[\,\gamma\,T^{WDD}\left(\vec{r}\right)\,\right]$, with a specific micrograph provided after each 1/8-portion of all scan positions has been treated. The required processing times are indicated as well, and the square rooted Fourier transform amplitude is given as an inset.}
            \label{fig:realtimeprocessNaClparticle}
        \end{figure*}
        
        To add to the last subsections, a third demonstration was conducted on NaCl particles. Details of their preparation are available in appendix 4. This ionic crystal structure was selected as a representative beam-sensitive object, as it is known to undergo rapid structural decomposition in the microscope, driven by both radiolytic and knock-on mechanisms \cite{Egerton2019}. Some prior work, reporting an advance in graphene encapsulation techniques, has demonstrated that even a brief exposure to the electron beam is sufficient to initiate irreversible degradation \cite{Lehnert2021}.
        
        The acquisition and reconstruction were performed using the same instrument and computing hardware as for the zeolite and SrTiO$_{\text{3}}$ specimens. Relevant experimental parameters include a scan grid of 2048$^{\text{2}}$ positions, a dwell time of 1 $\mu$s, a field of view of 13 nm, an acceleration voltage of 200 kV and a convergence half-angle of 18 mrad. Those choices lead to an overlap ratio of 98.72\%. The beam current and dose were respectively estimated to 1.2 pA and 1800 $e^-/\text{\AA}^2$. For reference, the PACBED pattern and an isolated CBED frame, reformed from the event-based dataset, are available in fig. \ref{fig:exppatternNaClparticle}. Processing parameters were also identical to those chosen in previous use cases.
        
        The results of GPRI-based WDD ptychography, including snapshots of the partially completed processing, are provided in figure \ref{fig:realtimeprocessNaClparticle}. For this new measurement, the total needed duration was 3.60 s, which implies an investment of 0.86 $\mu$s in the treatment of a single equivalent diffraction pattern, on average. Strikingly, this is even shorter than noted in subsections \ref{subsec:expdemozeolite} and \ref{subsec:expdemoSTO}, and demonstrates a prospect in performing reconstructions at sub-$\mu$s rates, which would allow matching the maximum reasonable acquisition speeds reported in ref. \cite{Jannis2022}. Of course, in that situation, the processing time also reflect the lower amount of electrons received per scan position, which further highlights the flexibility for real-space sampling, as explained prior.
        
        Concerning the imaged NaCl salt particle, it is noteworthy that its size was well beyond the normal range of validity for the thin specimen approximation \cite{Cowley1972,Gibson1994}, or for a good conservation of coherence \cite{Kwon2026,Mendis2026a}. While material thickness within the scanned area was not measured in this work, due to this not being critical for method demonstration purposes, a large-scale view of the nanocrystal tends to put its value close to 200 nm. This is explained in appendix 4.
        
        As a consequence, the results provided in fig. \ref{fig:realtimeprocessNaClparticle} were obtained in an highly unfavorable condition. Specifically, incoherent and diffuse scattering is then expected to play a major role \cite{Hall1965a,Loane1991,Wang1995,Muller2001,Mkhoyan2008,Beyer2020,Barthel2020,Barthel2021,Robert2022}, likely limiting the highest achievable resolution \cite{VanDyck2011,VanDyck2015}, and heavy nonlinearity of image contrast should occur as well \cite{Plamann1998,Yang2017,Gao2022,Clark2023,Leidl2025}.
        
        Even then, micrographs could still be obtained showing clear atomic features and providing information on the local lattice structure. Interestingly, the specific pattern of beam damage is also visible, corresponding to long-range information in the image spectrum, transferred alongside the crystal frequencies and highlighting beam-induced transformations.
        
        Like in subsection \ref{subsec:expdemoSTO}, this reconstruction of \\$Im\left[\,\gamma\,T^{WDD}\left(\vec{r}\right)\,\right]$ did not involve a computational correction of the residual aberrations in the optical system. It is nevertheless likely that the recording conditions, refined directly in-experiment, encompassed a probe focus able to partly compensate propagation-induced contrast redistribution effects \cite{Robert2021}. For instance, by placing it at a specific depth within the bulk, as was proposed in ref. \cite{Clark2023,Gao2024}. As such, the present work represents a direct experimental demonstration of this prospective strategy, and suggests its applicability even in extreme situations.
    
    \subsection{Large fields of view - NaCl bulk crystal}
    \vspace{3pt}
        \label{subsec:expdemoNaClbulk}
        
        \begin{figure}
            \centering
            \includegraphics[width=0.9\columnwidth]{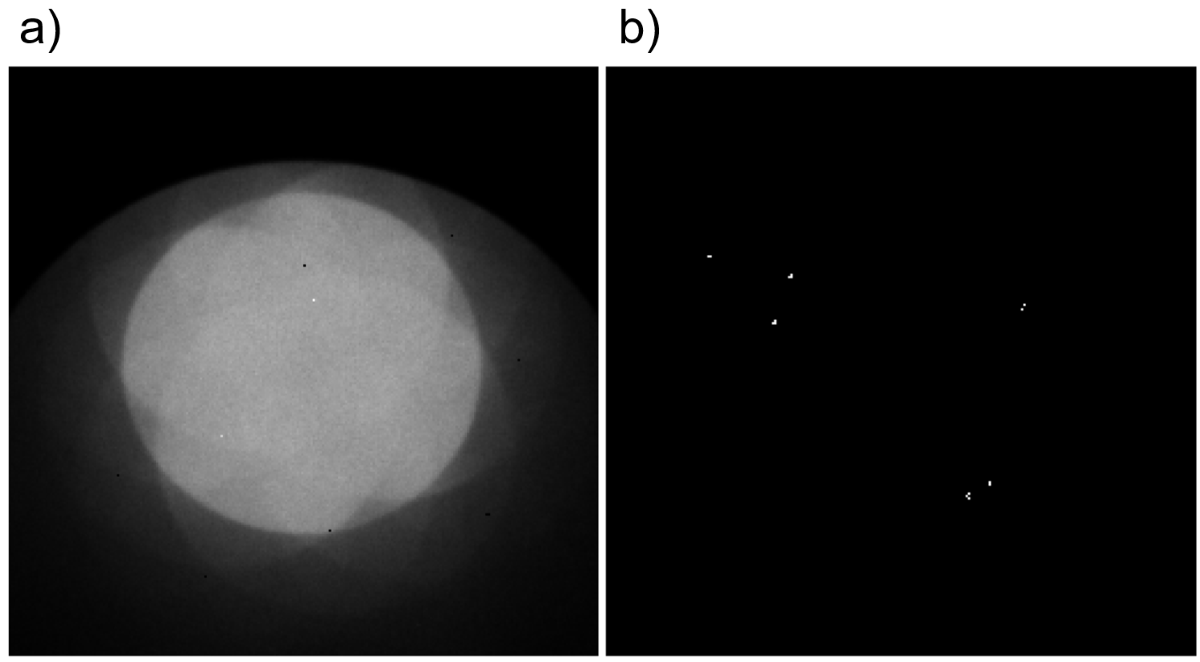}
            \caption{Reformation of 2D diffraction frames from the experimental event-based data, recovered from the NaCl bulk crystal specimen. a) PACBED pattern, b) single CBED pattern.}
            \label{fig:exppatternNaClbulk}
        \end{figure}
        
        \begin{figure*}
            \centering
            \includegraphics[width=1.0\textwidth]{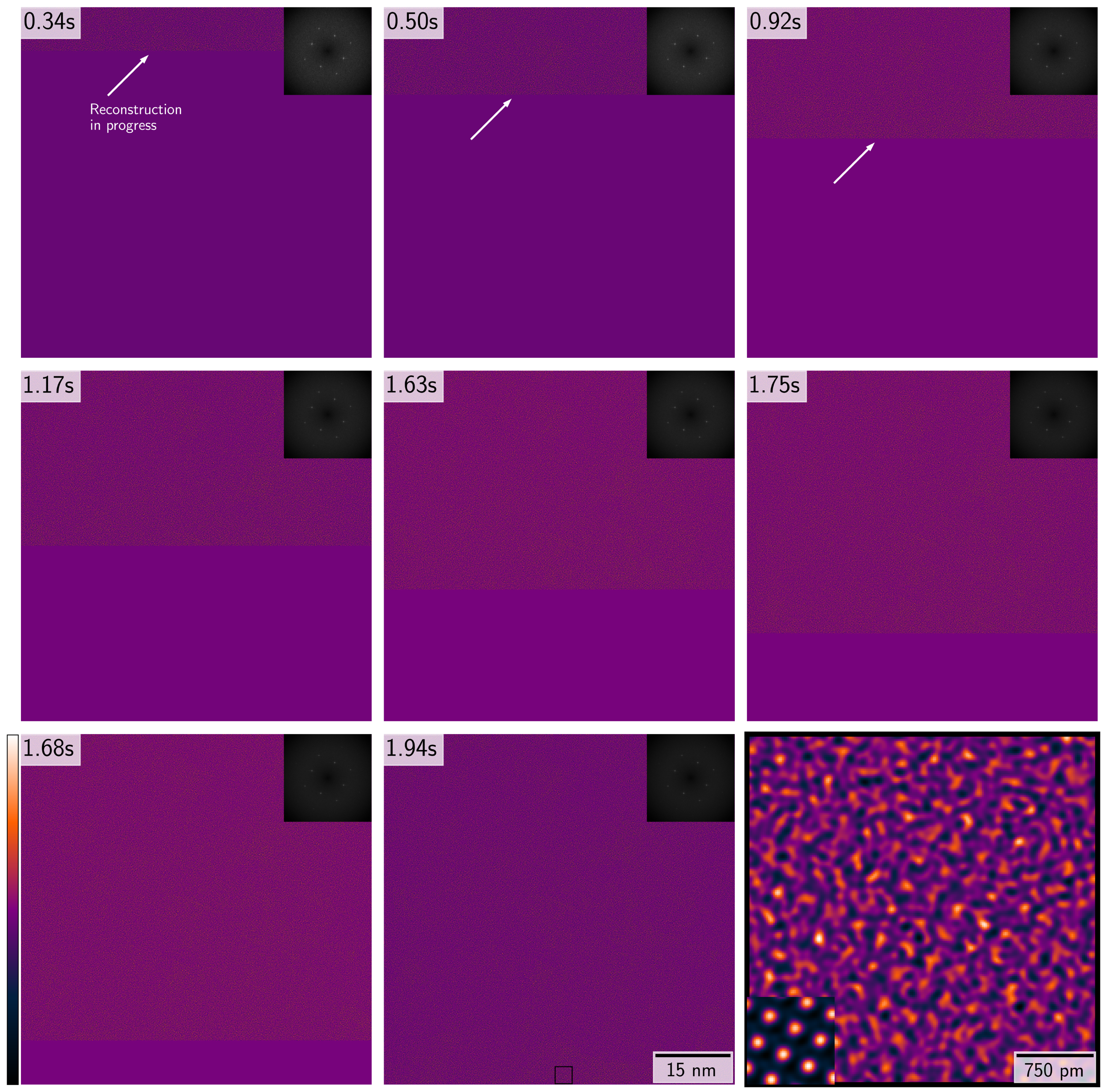}
            \caption{Result of an event-driven WDD reconstruction, done via the GPRI framework, applied on the NaCl bulk crystal specimen. Illumination parameters include an acceleration voltage of 200 kV, a convergence semi-angle of 19 mrad, an approximate dose of 100 $e^-/\text{\AA}^2$ and a scan step of about 34 pm. The figure depicts the quantity $Im\left[\,\gamma\,T^{WDD}\left(\vec{r}\right)\,\right]$, with a specific micrograph provided after each 1/8-portion of all scan positions has been treated. The required processing times are indicated as well, and the square rooted Fourier transform amplitude is given as an inset. The last panel (bottom right) shows a square cut region of the fully completed reconstruction, indicated there in a black rectangle. The result of a template matching procedure, based on the whole image, is added with the same spatial scale.}
            \label{fig:realtimeprocessNaClbulk}
        \end{figure*}
        
        With some demonstrations of fast processing and high resolution having been done over the last three subsections, it remains to illustrate the capacity for large fields of view, as is enabled by the event-driven acquisition/reconstruction pipeline. To this end, one last experiment was conducted on an NaCl bulk specimen, oriented along the [100] zone axis. From the fabrication process, the thickness of this material is expected to be close to 30 nm, but was not accurately measured due to it not being fundamentally needed in this work. Some additional details are available in appendix 4. The acquisition was done with the same instrument as described prior, and the GPRI processing was performed using the same computing hardware as well.
        
        The recording involved a scan grid of 2048$^{\text{2}}$ positions and a dwell time of 1 $\mu$s. The acceleration voltage was set to 200 kV and the convergence half-angle to 19 mrad. Compared to previous cases, the scan interval was given a relatively high value, leading to a field of view of 70 nm and an overlap ratio of 67.4\%, much more typical in focused probe-based ptychography \cite{Bunk2008}. To illustrate the area covered by the electron beam on the Timepix3 chip, a PACBED frame and an isolated diffraction pattern are provided in figure \ref{fig:exppatternNaClbulk}.
        
        Like in subsection \ref{subsec:expdemoSTO}, the HAADF detector was left inserted, as this was more practical to conduct the acquisition under those relatively challenging experimental conditions. Consequently, the measured values of beam current and dose, respectively close to 1.0 pA and 100 $e^-/\text{\AA}^2$, are expected to be slightly underestimated. Even then, they remain indicative of a very low amount of electrons to be invested in this measurement, especially given the targeted resolution \cite{Egerton2007,LalandecRobert2025}. Under those parameters, the long-range crystallinity of the specimen could be conserved, even without relying on prior specimen encapsulation \cite{Lehnert2021}. That, by itself, is a promising demonstration of the dose-efficiency of analytical ptychography, which also presents no risk of divergence in the phase retrieval process \cite{Katvotnik2013,Dearg2025,Chennit2025}.
        
        Fig. \ref{fig:realtimeprocessNaClbulk} displays the partial results of the GPRI procedure, for the dataset described above. The extraction of $\gamma\,T^{WDD}\left(\vec{r}\right)$ was done under the same processing choices as for the other use cases, and the figure follows the same general format as well, with 1/8-portions of the scan grid taken as reference points. A close look at the micrographs indicates a conservation of atomic visibility, though of course with the expected degree of noise \cite{OLeary2020,OLeary2021,LalandecRobert2025,Dearg2025}, essentially over the whole field of view. In the bottom right panel, a small isolated area of the full reconstruction is depicted, as an example. There, an additional inset is also provided, displaying the result of a template matching procedure and thus demonstrating a good recovery of the [100]-projected lattice structure in NaCl.
        
        In summary, for this rather wide illuminated surface of 70 nm by 70 nm, an accordingly large micrograph is obtained, with the expected crystalline pattern retrieved under low-dose conditions. For comparison, images produced in ref. \cite{Chen2020} reached a scale of 30 nm by 30 nm, while employing a defocus of 55 nm, an acceleration voltage of 80 kV and a convergence half-angle of 21.4 mrad. Here, just like for the other specimens, the probe remained well-focused and no computational correction of aberrations was attempted.
        
        Crucially, the total processing time was 1.94 s, leading to an average duration of 0.46 $\mu$s for the treatment of a single equivalent CBED pattern, which is the shortest reported so far in this work. This remarkable result is representative of the capabilities of the GPRI framework when optimal conditions are brought together, i.e. a reasonable value of overlap ratio and a low amount of counts, which would otherwise lead to a sparse frame-based dataset. As such, the images in fig. \ref{fig:realtimeprocessNaClbulk} should be generally regarded as an example of what can be made possible in routine microscopy work, in the near future.

\section{Discussion}
    
    \subsection{Introduction of real-time performance and large fields of view in ptychography}
    \vspace{3pt}
        
        The typically small fields of view achieved in MR-STEM, and thus in electron ptychography, can be related to the limited stability of the employed microscopes. In particular, it implies a practical upper bound of a few minutes for the total acquisition time. This then leads to a maximum number of scan points for a certain probe position dwell time, itself being fixed by the detector. In that respect, across existing experimental setups, conventional frame-based DED are currently dominant, with best obtainable recording rates in the order of 10$^5$ frames per second \cite{Zambon2023,Ercius2024}. As such, while this still represents a strong improvement from earlier models, whose highest frequencies were below 10 kHz \cite{Ballabriga2011,Ryll2016,Tate2016}, the feasible number of scan positions $N_s$ is still restricted to values well under 1024$^2$ in most practical use cases.
        
        In that context, a major interest of employing an EDD \cite{Poikela2014,Frojdh2015,Llopart2022} is to enable scanning the probe across a wider area \cite{Jannis2022}, and thus enhance measurement statistics. Nevertheless, having acquired electrons in a scan grid containing more than 1024$^2$ points, each leading to an equivalent 2D diffraction pattern, can make the application of conventional analytical ptychography workflows \cite{Li2014,Pennycook2015} overly time- and memory-intensive. As was already noted in subsections \ref{subsec:datapartitioning} and \ref{subsec:numericalaspects}, this is a fundamental consequence of their associated dense data representation, leading to a non-linearly increasing $N_s$-wise numerical complexity.
        
        Concurrently, detector technologies are still improving at a rapid pace, an example of which being the Timepix4 chip \cite{Llopart2022}, which was recently tested in a TEM setup \cite{Dimova2025,Ding2026}. Whereas the practical time resolution of this device is likely to be limited by in-sensor stochastic scattering \cite{Auad2024}, its achievable data output is still expected to allow major improvements from prior EDD models. Specifically, whereas the Timepix3 chip can accommodate a beam current of a few pA \cite{Jannis2022,Annys2025} in generating an event stream with no saturation effect, the high bandwidth of a Timepix4-based detector could upgrade this capability beyond 50 pA, thus enabling an even wider range of acquisition speeds.
        
        While this ever-evolving experimental capability keeps unfolding, it becomes increasingly clear that a naive adaptation of existing direct phase retrieval approaches to in-acquisition processing, as was the original solution chosen in ref. \cite{Strauch2021,Bangun2023}, will be insufficient in the long run. At the very least, persisting in that direction would entail excessive hardware requirements, thus making data transfer into a new bottleneck for those methods and introducing significantly higher cost-of-access for non-specialized users.
    
    \subsection{Interests of quantization-based phase retrieval}
    \vspace{3pt}
        \label{subsec:quantizationbasedphaseretrieval}
        
        Under the growing need for a ptychographic workflow with a performance high enough to keep up with state-of-the-art detector capacities, a quantization-based description, as is the foundation of GPRI, becomes most useful. Crucially, this permits a treatment of an arbitrarily small part of the complete scattering dataset, from the single count level to a few-$\vec{r}_s$ packet. All the typical time-intensive aspects of a WDD or SBI reconstruction are furthermore condensed within a unique library pre-calculation step. In particular, as shown in section \ref{sec:theorysection2}, the typical computation pathway of WDD, i.e. Fourier transforming the data and dividing it with a Wigner distribution along the phase space axes $\left[\,\vec{Q}\,;\,\vec{R}\,\right]$, is equivalently represented in the preparation and usage of $G^{WDD}_{\vec{q}_d}\left(\vec{r}\right)$.
        
        As a consequence, the GPRI framework offers essentially unlimited flexibility for parallelization, while being relatively easy to implement, at least in its most naive form. At the level of a single event, a measurement step simply consists in the $\vec{q}_d$-wise choice of a kernel-limited guide function from within the library, a scan position-dependent area selection within the wider reconstruction window, and an addition. This approach allows the removal of all calculation steps involving empty pixels within a sparse dataset, which would otherwise be unavoidable in a strict frame-based processing.
        
        Relying on an appropriate software architecture for in-acquisition data partitioning, a short few-counts treatment then permits a processing speed that can surpass the maximum event rate in an EDD. This was demonstrated for the typical range of beam currents used with a Timepix3 chip \cite{Annys2025}, thus offering a straightforward implementation for real-time imaging. The experimental results provided over section \ref{sec:experiments} further exemplify this capability, with total processing times in the order of a few seconds. Interestingly, the dependence of the computational cost on the number of counts, rather than on the number of scan points, also facilitates the treatment of large MR-STEM acquisitions, e.g. including 2048$^{\text{2}}$ positions. This then can be used to treat very densely sampled areas, and makes it more straightforward to achieve wide fields of view.
        
        Importantly, no pre-processing of the CBED patterns or count locations is needed, as the physical $\vec{q}_d$ vectors used in the calculation of the library are simply taken from a prior calibration of detector space. This is explained in more details in appendix 1.
        
        Moreover, as long as the detection coordinates are accessible directly upon reception by the EDD, no intermediary in-disk saving is needed. Immediate calculation can then be performed while optionally discarding the scattering data, thus permitting a rather important economy of storage space in routine experiments. Of course, GPRI imaging in general, whether it is used to recover a definitive result or simply for an initial real-time observation and microscope alignment step, does not preclude the recovery of complete diffraction patterns and e.g. a follow-up iterative optimization procedure.
        
        In general, GPRI offers an advantageous alternative for the majority of use cases in analytical ptychography. The only exceptions are the interactive post-acquisition refinement of the aberration function \cite{Yang2016,Li2025a}, which is based on the WPOA and requires the partially reformulated dataset $\tilde{J}_{\vec{Q}}\left(\vec{q}_d\right)$ to be fully available in-memory, and the usage of an excessively defocused probe \cite{Hue2010,Song2019}. This latter limit is a consequence of the $\vec{r}$-wise restriction of the library, which in principle reflects the area covered by the illumination. In particular, beyond a certain degree of aberration-induced expansion, the required kernel radii become too large to be manageable by GPU memories. For this reason, and also because a smaller kernel implies less pixel-wise updates, GPRI typically performs better for an acquisition based on a densely scanned, and well-focused, probe.
        
        Even then, it is worth noting that, under a reasonable value of defocus, the new methodology remains applicable. Initial verifications have demonstrated quite tolerable memory cost in experimental conditions reported for various recent use cases, e.g. in ref. \cite{Zhu2025,Harikrishnan2026,Byrne2026}, using a commercial computer. This indicates the concurrent usability of this recording geometry, increasingly relevant for its expected capacity to tune the contrast transfer for sub-$q_A$ spatial frequencies \cite{Dwyer2024,Varnavides2025,Ma2025,Ma2026,Bennemann2026}. In other words, the kernel size limitation remains far from strict, and it would quickly become irrelevant given some moderate upgrades of the computing hardware. Such a usage will be explored in further publications.
    
    \subsection{Comparison to iterative optimization methods}
    \vspace{3pt}
        
        The numerical simplicity of GPRI offers a substantial help for practical implementations, especially in contexts where calculation time and hardware cost are restricted. As such, this methodology promises high accessibility for non-specialized facilities, e.g. having no need for an assistance by high-performance computing \cite{Jones2022,Mukherjee2022,Wang2022c,Welborn2024}. Moreover, thanks to an inclusion within a simple interface \cite{Annys2025} and due to requiring very little input parameters, it can be employed with ease by non-experts. In the future, a highly simplified workflow could also be envisioned where libraries of guide functions are available at all times, locally and for a collection of standard illumination conditions. As such, this new solution would be straightforwardly enabled in routine scientific work, being largely aided by a live feedback permitting interactive modifications of illumination parameters.
        
        It is interesting to relate those various prospects to the current context of ptychography, within which iterative optimization methods are dominant. In general, this relative hegemony is justified by the existence of sophisticated solutions for e.g. the refinement of the probe model \cite{Thibault2009,Maiden2009,Thibault2013,Chen2020}, super-resolution \cite{Maiden2011,Humphry2012,Jiang2018} or 3D reconstructions based on a multislice propagation \cite{Maiden2012,Gao2017,Chen2021}. Iterative ptychography thus constitutes an elaborate computational imaging tool, capable of solving complex physical problems while accommodating some uncertainties in the illumination parameters and material thickness. In particular, accessing specimen frequencies beyond the diffraction limit, as has been an historical goal of the wider field \cite{Sayre1952,Gerchberg1974,Rodenburg1992}, is likely its most attractive feature.
        
        Nevertheless, concurrent to their qualities, those optimization approaches typically come with very long processing times, while sometimes presenting issues of convergence, especially in the low-dose case \cite{Katvotnik2013,Dearg2025,Chennit2025}. In particular, this latter issue often requires extensive parameter optimization \cite{Maiden2017,Bangun2022,Leidl2024,Maiden2024,Chennit2025}. Those algorithms are also generally difficult to implement and set up in practice due to their complexity and, somewhat paradoxically, the variety in their modalities.
        
        In fact, it could even be argued that, for a lot of potential applications, their various advantages may be superfluous. Specifically, a computational improvement of resolution beyond the normal performance of an instrument, though useful within specific investigations, is not systematically needed. A similar statement can be made with regards to aberration refinement. Even the ability to account for in-specimen propagation effects, typically inducing contrast non-linearity and modifications of frequency transfer \cite{Plamann1998,Yang2017,Gao2022,Clark2023,Leidl2025}, may find an alternative in direct phase retrieval methods, since a specific choice of probe focus can be sufficient to suppress the resulting spatial redistribution and ensure interpretability. This has been demonstrated in iCoM imaging \cite{Close2015,Addiego2020,Burger2020,Robert2021,Liang2023}, as well as for the SBI and WDD methods \cite{Clark2023,Gao2024}.
        
        Notably, the experimental results provided in this publication already confirm the applicability of analytical ptychography for very thick objects, granted some adjustments in illumination parameters. That approach, of course, will have to be further evaluated in the future and cannot replace a more rigorous solution, but this still shows that recovering interpretable images remain, at the very least, possible in such unfavorable conditions. Reconstructed transmission functions also clearly demonstrate the possibility to achieve satisfactory high-resolution results, with an aberration-corrected STEM instrument \cite{Batson2002,Erni2009,Morishita2018a}, even without a computational correction step.
        
        For those reasons, while it remains certain that iterative ptychography can lead to impressive results, the analytical variant clearly has the upper hand in terms of both accessibility and reproducibility. Consequently, and for all its other advantages, the GPRI framework can be a first-line option in most phase contrast imaging endeavours, even in cases where a more sophisticated optimization process is required later on. As such, this new methodology constitutes a complementary solution, able to fill a clear gap in the field.
    
    \subsection{Comparison to the optimum bright field method}
    \vspace{3pt}
        \label{subsec:comparisontoOBF}
        
        As has been hinted in the introduction, and illustrated by equation \ref{eq:continuousWDDGPRI}, a similarity can be found between the GPRI methodology and the OBF-STEM technique \cite{Ooe2021,Ooe2023,Ooe2024,Ooe2026}, at least when the latter is applied on the pixels of a DED. That is, both consist in a linear superposition of scan position-shifted functions, with each of those corresponding to one of the detector pixels and being weighted by its specific received intensity. This parallel nevertheless remains restricted to a superficial signal processing level, and two key differences should be pointed out.
        
        The first, and most fundamental, is the explicit usage of quantization within the scattered intensity, and not just from the practical aspect of observing sparsity in a low-dose acquisition \cite{OLeary2020,Pelz2022}. As was explained in details in subsection \ref{subsec:countwiseWDD1}, the theoretical foundation of GPRI is a modeling of incident electrons by Dirac delta-functions. This elementary tool allows a major simplification in the general analytical solution of ptychography for a multiplicative interaction model, i.e. the WDD method. While WDD imaging was originally described in ref. \cite{Rodenburg1992}, with some first tests reported in ref. \cite{McCallum1992,Chapman1996}, this simplification for an explicitly quantized radiation was not provided until now.
        
        On the other hand, the theoretical background of the OBF method is a simple integration of the far-field intensity within a detector-covered region, done while assuming a weakly scattering object. This additional approximation allows a simplification of equation \ref{eq:Wignerdistributions1}, to a formulation reported first in ref. \cite{Rodenburg1993}. Upon applying a specific image frequency-dependent integration scheme \cite{Yang2015b}, as used in an SBI processing \cite{Pennycook2015}, this then permits extracting the specimen-induced phase shift, under a known phase contrast transfer function (PCTF).
        
        More generally, the WPOA also allows expressing a scanned detector signal as a real-space convolution between the phase shift map and a function that depends on the shape and location of the integration region in reciprocal space, as well as on the illumination conditions. While such a formalism has been long known \cite{Rose1974,Rose1977,Dekkers1978}, the proposal of using multiple of those STEM signals for a combined deconvolutive process, where the detector/illumination contribution is removed by using specific kernels, was first made in ref. \cite{Ooe2021}. An initial experimental demonstration, employing a segmented DPC detector \cite{Lohr2012} and a frequency space-based convolution procedure, was later provided in ref. \cite{Ooe2023}. Furthermore, an application on MR-STEM data was reported in ref. \cite{Ooe2024}, and the prospect of employing a real-space convolution process for faster imaging in experimental cases, ensured by cropping the detector pixel-specific kernels, was discussed in ref. \cite{Ooe2023,Ooe2026}.
        
        One of the other reported interests of OBF-STEM is the possibility it offers for an informed noise normalization strategy \cite{Seki2018}. This approach for an improvement of the signal-to-noise ratio has been shown to be applicable to SBI imaging \cite{Seki2018,OLeary2021} as well, since it shares the same theoretical foundation. Moreover, it could in principle be extended to a GPRI-based form of the SBI method, as derived in appendix 2. In practice however, the WPOA is typically unfulfilled in high-resolution electron microscopy \cite{Gibson1994}, even for light matter specimens \cite{Vulovic2014,LalandecRobert2025}, which sheds doubt on the validity of the assumed noise distribution in the object spectrum. That arguably remains true even when propagation, within an extended thickness of weakly scattering matter, is considered \cite{Ooe2024,Leidl2025}.
        
        The second distinctive aspect of GPRI, compared to the OBF method, is its ability to directly treat isolated counts received at arbitrary scattering vector coordinates, as shown explicitly by equations \ref{eq:multicountWDDGPRI1} and \ref{eq:multicountWDDGPRI2}. Whereas the new framework is fully applicable to densely represented CBED patterns, its main relevance pertains to a usage for single detection events, e.g. in a data stream obtained from an EDD. This is an inherent conceptual difference from other methods. Finally, the formalism of GPRI is also interesting for predictions of contrast formation mechanisms, as illustrated in subsection \ref{subsec:extractableinformation}.
    
    \subsection{Prospect for a new state-of-the-art and cross-discipline imaging solution}
    \vspace{3pt}
        
        Rather than just an enhanced implementation of existing computational imaging methods, the GPRI framework constitutes a key element in establishing a new measurement paradigm, relevant across various scientific fields and usable in routine experiments. This follows a core notion of ptychography, in that it is generalizable to any coherent scattering setup as long as a redundancy condition can be fulfilled within the acquisition \cite{Bunk2008}.
        
        Hence, while this quantization-based methodology was initially developed for STEM-based phase retrieval, it is also appropriate e.g. in an X-ray diffraction system or with an optical microscope \cite{Wang2025a}. Such an adaptation would only require modifications of the functions $A\left(\vec{q}_0\right)$ and $\chi\left(\vec{q}_0\right)$ in equation \ref{eq:Wignerdistributions2}, reflecting differences in focusing optics. Consequently, a wide range of research facilities, including large-scale instruments such as a synchrotron source, could profit from it, likely with no new hardware.
        
        In that context, GPRI-enabled phase contrast can prove useful for a variety of currently developing topics, including e.g. biological matter \cite{Shahmoradian2017,Zhou2020,Wang2023b,Kucukoglu2024}, metal-organic frameworks \cite{Li2025}, battery materials \cite{Kunze2025,Sun2025}, magnetic textures \cite{Donnely2016,Chen2022,Mendoza2025,Butcher2025b} and ferroelectric crystals \cite{Harikrishnan2025,Butcher2025}. Specifically, it will permit setting up new experimental pipelines to conduct extensive recording series, with immediately available reconstruction results. The low numerical restrictions, with no possibility of convergence issues, will furthermore facilitate the usage of low radiation doses, down to few-counts sparsity \cite{OLeary2020}. Thanks to this major upgrade in measurement statistics and cost-of-access, breakthroughs may then be achieved across various fields, while remedying typical challenges of reproducibility.

\section*{Conclusion}
    
    The GPRI framework \cite{LalandecRobert2025a} offers a new paradigm for ptychography, and constitutes a step forward in the wider field of computational imaging. In particular, its underlying quantization-based description represents an intrinsic revision of the manner in which an acquisition of diffraction patterns is considered.
    
    In this publication, a formulation for the successive treatment of received detection events was introduced. Specifically, their individual contributions take the form of pre-calculated kernel-limited guide functions, only dependent on instrumental parameters. This essential notion could be used to demonstrate a novel implementation of the Wigner distribution deconvolution method. In parallel, this was also done for the SBI and iCoM approaches, as shown in appendix 2.
    
    Additionally, a verification was conducted using simulated diffraction patterns. There, a result showing identical dose-efficiency and performance to that of a conventional WDD implementation was obtained, with high benefits in terms of calculation time and memory requirements. A comparable application using the SFPA approach \cite{LalandecRobert2025} was provided in appendix 3. This was pursued with a demonstration on experimental event-driven data, encompassing a treatment of single patterns with sub-$\mu$s durations, very large fields of view and atomic resolution. Those results lead the way for a future integration in routine measurements.
    
    Within following works, the usage of GPRI-based phase contrast will be explored in cutting-edge applications. Crucially, its combination with rapidly improving detector technologies is expected to induce changes in the common practices of the field. That is because this framework complements existing iterative algorithms, filling a major methodological gap, and provides real-time measurement capacities with reproducible behavior at very low radiation doses. Given the increasing importance of ptychographic techniques across a variety of research fields, GPRI is foreseen to have an impact far beyond the scope of electron microscopy.

\section*{Appendixes}
    
    \subsection*{A.1. Details of the library calculation procedure}
    \vspace{3pt}
        
        One important aspect of the GPRI framework is its reliance on the pre-calculated library $G^{WDD}_{{\vec{q}_d}}\left(\vec{r}\right)$. That is, one specific kernel-limited guide function for each considered scattering vector in the far-field, thus implying dimensions $K_x \, K_y \, M_x \, M_y$ for the resulting variable, where $K_{x,\,y}$ represents the amount of pixels along the axes of a square detector. Generating this distribution is done in two general steps and involves a combination of FFT and DFT.
        
        First, the Wigner distribution $\Gamma\left(\vec{Q}\,;\,\vec{R}\right)$ has to be explicitly formed in-memory. This is done through equation \ref{eq:Wignerdistributions2}, and thus by introducing two distinct sets of numerically defined dimensions: a primary grid $\vec{Q}\,/\,\vec{r}$ and an auxiliary grid $\vec{q}_0\,/\,\vec{R}$. The process consists in calculating 2D aperture overlap profiles, including an optional aberration function, within $\vec{q}_0$-space. For each of those, a simple iFFT is then used to reach the $\vec{R}$-axes. This is done separately for all considered shifting frequency $\vec{Q}$.
        
        Importantly, as opposed to the conventional WDD workflow, the $\vec{R}$ dimensions do not depend directly on the native detector space. In practice, their grid simply has to be prepared such that its pixel size, relating to the maximum considered $\vec{q}_0$, remains small enough to accommodate a certain scattering vector cutoff, through which the user can decide to use electrons scattered e.g. up to a certain multiple of $q_A$, limited to the detector-covered $\vec{q}_d$-range. This will be further clarified in the rest of this appendix.
        
        As an additional condition, the maximum value of $\parallel\vec{R}\parallel$ is required to be large enough to avoid aliasing, upon inverse Fourier transforming the aperture overlap profile. Hence, the area needed for this auxiliary space, in the following referred to as a calculation kernel, as opposed to the previously mentioned for-processing kernel, is dependent on the illumination conditions. In practice, it is given a value of e.g. 8 to 16 times the Abbe criterion $\delta r_{Abbe} = 0.5/q_A$ to ensure robustness in the final results. As an additional option, a multiplication by a 2D Hann window along $\vec{R}$ may be done as well. As was explained in subsection \ref{subsec:phasespacedeconv}, this also has implications with regards to detector pixel size, represented by a certain MTF opening.
        
        With regards to the $\vec{Q}\,/\,\vec{r}$-coordinates, it is important to note that they, similarly, do not directly relate to the wider window to be used in the follow-up reconstruction. For the preparation of this primary calculation grid, the maximum value of $\parallel\vec{r}\parallel$ has to be chosen such that a following $\vec{Q}$-to-$\vec{r}$ iFFT, whose need will be explained in the following, also does not lead to an aliasing effect. This is normally solved by choosing a calculation kernel size equal to the one used in the auxiliary real-space grid $\vec{R}$.
        
        Concerning the pixel size along the $\vec{r}$-grid, which corresponds to the maximum introduced $\vec{Q}$, it is required to match that of the wider reconstruction window. As a first condition, this spatial dimension has to satisfy the user-defined frequency limit of the measurement, itself with an upper bound of $2\,q_A$ \cite{Nellist1994}, as was mentioned in subsection \ref{subsec:phasespacedeconv}.
        
        Additionally, the ratio of the experimental scan step over this real-space pixel size normally needs to be an integer. This ensures that physical $\vec{r}_s$-locations, without considering a possible scanning error, correspond to round numerical coordinates in the reconstruction grid, and that scan position-shifting is done by unambiguous selection of in-kernel pixels. If this condition is not fulfilled, an FFT-based shifting or real-space interpolation step becomes necessary, which defeats the purpose of the wider algorithm by increasing the duration of its individual calculation steps. As an alternative, the scan positions could simply be approximated to then-decorrelated reconstruction pixels, themselves with a diffraction-limited sampling, at a small cost in measurement accuracy.
        
        Importantly, the four-dimensional $\left[\,\vec{Q}\,;\,\vec{R}\,\right]$-grid, within which the initial generation of $\Gamma\left(\vec{Q}\,;\,\vec{R}\right)$ is performed, is typically found to be very compact, and can be accommodated by the memory of a commercial GPU. As explained above, this is thanks to the independence of those dimensions from both the detector axes and the reconstruction window, which only leaves the size of the probe to be considered. Such illumination-dependent compactness is another specificity of the GPRI-based form of the WDD method, compared to the conventional approach. That is, the phase space region required for calculation is spatially localized, i.e. at the level of a single scan position-specific interaction. In comparison, the conventional approach enforces a representation of the illumination that extends to the full four-dimensional space covered by the specimen function $\Upsilon\left(\vec{Q}\,;\,\vec{R}\right)$ to be extracted, regardless of the actual probe size.
        
        Once the Wigner distribution is available in-memory, the second step of the library calculation process consists in the generation of the guide functions themselves, based on equation \ref{eq:WDDguidefunction}. To this end, $\Gamma\left(\vec{Q}\,;\,\vec{R}\right)$ is optionally multiplied by a known MTF $M\left(\vec{R}\right)$, and inverted through a Wiener filter \cite{Bates1986}, given a user-defined parameter $\epsilon$. Following this, a frequency-wise $\tilde{G}^{WDD}_{{\vec{q}_d}}\left(\vec{Q}\right)$ is extracted for each considered scattering vector $\vec{q}_d$ separately. Moreover, and as mentioned in subsection \ref{subsec:phasespacedeconv} for the conventional phase space-wise deconvolution approach, those detector coordinates are calibrated prior \cite{Robert2021} and thus include a direct measurement of the scan raster-to-camera rotation effect, as well as possible projection system-induced elliptical distortions \cite{Capitani2006}.
        
        This explicit detector space calibration makes it possible to perform the latter $\vec{R}$-to-$\vec{q}_d$ information transfer through a DFT, rather than an FFT. More precisely, a series of $\vec{q}_d$-specific calculations is conducted, each involving the explicit multiplication of the inverted Wigner distribution by a 2D plane wave term $e^{-i 2\pi \vec{q}_d\cdot\vec{R}}$ and a summation, both done along the auxiliary space $\vec{R}$. As a side-note, this requires appropriate Fourier normalization, here chosen orthonormal like in ref. \cite{LalandecRobert2025}. Another aspect to consider is the user-defined $\vec{q}_d$-cutoff, i.e. a scattering vector modulus beyond which the guide function is left undefined. As mentioned above, this parameter determines the pixel size within $\vec{R}$-space. It also allows the removal of a possible reconstruction artefact, due to the miscentering of the CBED pattern over the detector. Finally, iFFT are applied to the resulting $\vec{Q}$-dependent distributions, thus reaching back the assigned primary grid $\vec{r}$, limited to the initially chosen calculation kernel size.
        
        As a final step, those $\vec{q}_d$-specific guide functions are truncated to a second smaller area. This is the previously mentioned for-processing kernel, which may be assigned a radius of e.g. 4 to 8 times $\delta r_{Abbe}$, having to reflect the size of the probe, but this time without a risk of aliasing. Another Hann window is optionally applied along the reduced $\vec{r}$ axes to avoid cutoff artefacts and to correct for unwanted modifications of frequency transfer \cite{Yu2022a,Ooe2026}. Crucially, thanks to this spatial restriction, a very manageable memory requirement is ensured for both the storage and usage of the library.
    
    \subsection*{A.2. GPRI-based formulation of SBI and iCoM}
    \vspace{3pt}
        
        The second existing analytical ptychography method, here referred to as SBI and sometimes called single sideband (SSB) \cite{Yang2017} reconstruction, requires the supplementary assumption of a weakly scattering specimen, i.e. the previously mentioned WPOA. Moreover, it exists in two forms: a deconvolutive modality \cite{Yang2016a,LalandecRobert2025} (SBI-D) and a summative one \cite{Pennycook2015,Yang2015b} (SBI-S). The limitations of those approaches in terms of frequency transfer, in particular due to the assumed phase contrast transfer function \cite{Yang2015b} and the unrealistic fulfillment condition of the WPOA, have been discussed e.g. in ref. \cite{LalandecRobert2025,Leidl2025}.
        
        As a further alternative STEM-based phase retrieval method, iCoM imaging consists in the Fourier integration \cite{Lazic2016,Yu2022a} of the average momentum transfer \cite{Muller2014} vector map, i.e. the center of mass of the scan position-wise CBED pattern. This measurement approach is affected by a supplementary optical transfer function (OTF) effect \cite{Black1957}, which tends to suppress higher spatial frequencies \cite{Yucelen2018}.
        
        The concept of GPRI, initially demonstrated for WDD in this publication, can be extended to the SBI and iCoM methods. Specifically, starting from the description provided in ref. \cite{LalandecRobert2025} and using equations \ref{eq:multicountdataandphasespacedata}, it is possible to derive
        \begin{equation}
            \varphi^{SBI}\left(\vec{r}\right) \, \propto \, \sum\limits_{\vec{r}_s} \sum\limits_{\vec{q}_d} \, I^{exp}_{\vec{r}_s}\left(\vec{q_d}\right) \, G^{SBI}_{\vec{q}_d}\left(\vec{r}-\vec{r}_s\right) \quad,
        \end{equation}
        and
        \begin{equation}
            \label{eq:continuousiCoMGPRI}
            \varphi^{iCoM}\left(\vec{r}\right) \, \propto \, \sum\limits_{\vec{r}_s} \sum\limits_{\vec{q}_d} \,  I^{exp}_{\vec{r}_s}\left(\vec{q_d}\right) \, G^{iCoM}_{\vec{q}_d}\left(\vec{r}-\vec{r}_s\right) \quad,
        \end{equation}
        here using the same continuous description of the experimental intensity as in formula \ref{eq:continuousWDDGPRI}. This can be rewritten count-wise, similarly to equations \ref{eq:multicountWDDGPRI1} and \ref{eq:multicountWDDGPRI2}. The distributions $G^{SBI}_{\vec{q}_d}\left(\vec{r}\right)$ and $G^{iCoM}_{\vec{q}_d}\left(\vec{r}\right)$ are two sets of guide functions for SBI and iCoM. Notably, while the previously introduced $G^{WDD}_{\vec{q}_d}\left(\vec{r}\right)$ is complex-numbered, those new libraries are real quantities, as they aim to provide measurements of the phase shift map $\varphi\left(\vec{r}\right)$ directly.
        
        In SBI-D, the derived $\vec{Q}$-space expression of the library is
        \begin{equation}
            \label{eq:SBIDguidefunction}
            \tilde{G}^{SBI}_{\vec{q}_d}\left(\vec{Q}\right) \, = \, \frac{1}{i \, \sum\limits_{\vec{q}_d^{\,\,\,'}} A^2\left(\vec{q}_d^{\,\,\,'}\right)} \, \frac{ \tilde{\omega}^*\left(\vec{Q}\,;\,\vec{q}_d\right) }{ \epsilon \, + \, \mid\tilde{\omega}\left(\vec{Q}\,;\,\vec{q}_d\right)\mid^2 } \quad,
        \end{equation}
        with
        \begin{equation}
            \omega\left(\vec{Q}\,;\,\vec{R}\right) \, = M\left(\vec{R}\right) \, \Gamma\left(\vec{Q}\,;\,\vec{R}\right) \, \left( e^{i2\pi\vec{Q}\cdot\vec{R}} - 1 \right) \quad.
        \end{equation}
        This set of guide functions thus takes the form of the Fourier transformed, and then inverted, illumination Wigner distribution, with introduction of a $\vec{Q}$-wise shifting term $\left( e^{i2\pi\vec{Q}\cdot\vec{R}} - 1 \right)$. This is a consequence of the sideband-based description \cite{Rodenburg1993}, which becomes effective in the framework of the WPOA. Like in the WDD case, this encompasses an explicit role for the MTF $M\left(\vec{r}_d\right)$ of the detector. When this role is neglected, the SBI-S approach becomes applicable and the retrieval of the projected potential may be performed by frequency-wise summations of double overlap features in the scattering data, as explained e.g. in ref. \cite{Pennycook2015}. The library of guide functions then takes the form
        \begin{equation}
            \label{eq:SBISguidefunction}
            \tilde{G}^{SBI}_{\vec{q}_d}\left(\vec{r}\right) \, = \, \frac{1}{i} \, \left( \, \beta^{+}\left(\vec{Q}\,;\,\vec{q}_d\right) \, - \, \beta^{-}\left(\vec{Q}\,;\,\vec{q}_d\right) \, \right) \quad,
        \end{equation}
        with
        \begin{equation}
        \begin{split}
            \beta^{+}\left(\vec{Q}\,;\,\vec{q}_d\right) \, & = \, \frac{A\left(\vec{q}_d\right) \, A\left(\vec{q}_d-\vec{Q}\right) \, \left( 1 - A\left(\vec{q}_d+\vec{Q}\right) \right)}{e^{-i\left( \chi\left(\vec{q}_0-\vec{Q}\right) - \chi\left(\vec{q}_0\right) \right)}} \\
            \beta^{-}\left(\vec{Q}\,;\,\vec{q}_d\right) \, & = \, \frac{A\left(\vec{q}_d\right) \, A\left(\vec{q}_d+\vec{Q}\right) \, \left( 1 - A\left(\vec{q}_d-\vec{Q}\right) \right)}{e^{-i\left( \chi\left(\vec{q}_0\right) - \chi\left(\vec{q}_0+\vec{Q}\right) \right)}} \quad.
        \end{split}
        \end{equation}
        As such, the functions $\beta^{\pm}\left(\vec{Q}\,;\,\vec{q}_d\right)$ take the role of complex-numbered $\vec{Q}$-dependent virtual detectors, reflecting the conventional SSB workflow \cite{Pennycook2015,Yang2015b}. Importantly, in both equations \ref{eq:SBIDguidefunction} and \ref{eq:SBISguidefunction}, $G^{SBI}_{\vec{q}_d}\left(\vec{r}\right)$ is equal to zero for $\parallel\vec{q}_d\parallel \geq q_A$. This can be understood straightforwardly as a secondary consequence of the WPOA \cite{Rodenburg1993,Pennycook2015}, in that a weak phase object is not expected to scatter any electrons toward the dark field. As an additional remark, equation \ref{eq:SBISguidefunction} is also what the typical expression for OBF-STEM \cite{Ooe2021} would converge to, when considering infinitely small, and arbitrarily located, pixels as detectors and while excluding the WPOA-based noise normalization term \cite{Seki2018,OLeary2021}.
        
        Continuing, $G^{iCoM}_{\vec{q}_d}\left(\vec{r}\right)$ is straightforwardly derived as
        \begin{equation}
            \label{eq:iCoMguidefunction1}
            \tilde{G}^{iCoM}_{\vec{q}_d}\left(\vec{Q}\right) \, = \, \vec{q}_d \, \cdot \, \tilde{\vec{\nu}}\left(\vec{Q}\right) \quad,
        \end{equation}
        with $\vec{\nu}\left(\vec{r}\right)$ a vector field given in Fourier space by
        \begin{equation}
            \label{eq:iCoMguidefunction2}
            \tilde{\vec{\nu}}\left(\vec{Q}\right) \, = \, 
            \begin{cases}
                \, \vec{0} & \,\,\text{if } \parallel\vec{Q}\parallel \, = \, 0 \, \text{nm}^{-1} \\
                \, \frac{1}{i} \, \frac{ \vec{Q} }{ \parallel\vec{Q}\parallel^2 } & \,\,\text{otherwise}
            \end{cases}
            \quad.
        \end{equation}
        In those expression, $\vec{\nu}\left(\vec{r}\right)$ takes the role of an integrating kernel, and the dependence of the result on the scattering vector is linear. As an alternative to introducing two cases in equation \ref{eq:iCoMguidefunction2}, a small number $\epsilon$ may be added to the denominator to avoid divisions by zero, as was done e.g. in ref. \cite{LalandecRobert2025}. Formalized as a post-acquisition processing, GPRI-based iCoM is thus equivalent to
        \begin{equation}
            \varphi^{iCoM}\left(\vec{r}\right) \, \propto \, \sum\limits_{\vec{r}_s} \, \langle\,\vec{q}\,\,\rangle_{\vec{r}_s} \, \cdot \, \vec{\nu}\left(\vec{r}-\vec{r}_s\right) \quad,
        \end{equation}
        with a prior calculation of
        \begin{equation}
            \langle\,\vec{q}\,\,\rangle_{\vec{r}_s} \, = \, \sum\limits_{\vec{q}_d} \, \vec{q}_d \, I^{exp}_{\vec{r}_s}\left(\vec{q_d}\right) \quad,
        \end{equation}
        done within a given selection of $\vec{r}_s$-coordinates. Importantly, this formulation is the same as used for the real-time iCoM (riCoM) approach reported in ref. \cite{Yu2022a}. In comparison, in the GPRI variant, the scan position-wise contribution is computed count-by-count using the library, instead of first calculating the local center of mass vector.
    
    \subsection*{A.3. Comparative SFPA result for the apoferritin case}
    \vspace{3pt}
        
        \begin{figure}
            \centering
            \includegraphics[width=0.8\columnwidth]{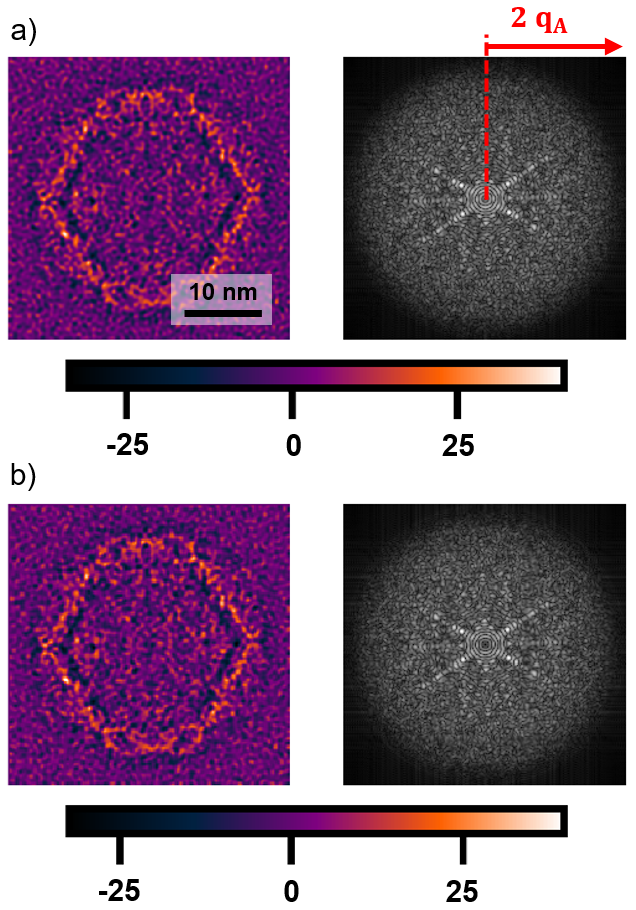}
            \caption{Comparison of the image contrast obtained in two implementations of a WDD reconstruction, specifically a) GPRI and b) SFPA. A Poisson expectation value of $N_{e^-}=256$ was employed in both cases. The figure depicts the position-dependent measurement of the projected potential $\mu^{WDD}\left(\vec{r}\right)$, alongside the square root of its Fourier transform's amplitude $\sqrt{\mid\tilde{\mu}^{WDD}\left(\vec{Q}\right)\mid}$. The colorbars are in V$\cdot$nm.}
            \label{fig:SFPAandGPRI}
        \end{figure}
        
        As a complement to subsection \ref{subsec:testonsim}, it is relevant to conduct a comparison of the image contrast obtained under the GPRI-based WDD reconstruction with that of another implementation. To that end, a projected potential map $\mu^{WDD}\left(\vec{r}\right)$ was extracted using the SFPA solution \cite{LalandecRobert2025}, from the same simulated CBED patterns as used in fig. \ref{fig:demonstration}. A Poisson expectation value $N_{e^-}=256$ was chosen, leading to a dose of about 41.5 $e^-/\text{\AA}^2$.
        
        The result, available in figure \ref{fig:SFPAandGPRI}.b, is essentially identical to the GPRI measurement, shown in fig. \ref{fig:SFPAandGPRI}.a for reference. This is of course expected, given that the processing approaches are physically equivalent, including in terms of their frequency transfer capacities and dose-efficiency \cite{OLeary2021,LalandecRobert2025}. The only notable difference is the pixel size in the reconstruction windows, chosen with specific and possibly differing requirements, as explained in subsection \ref{subsec:numericalaspects} and in appendix 1.
    
    \subsection*{A.4. Preparation of the NaCl specimens}
    \vspace{3pt}
        
        \begin{figure}
            \centering
            \includegraphics[width=1.0\columnwidth]{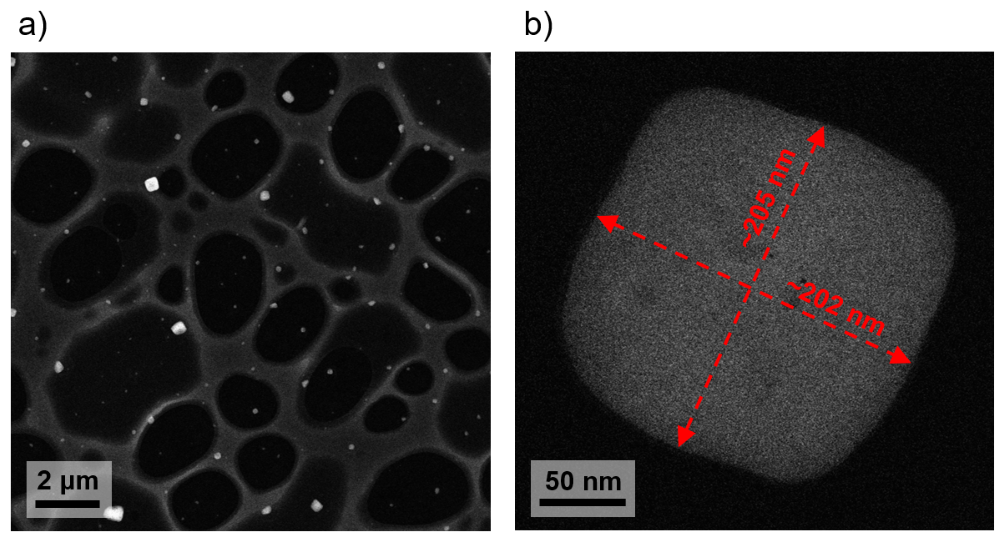}
            \caption{STEM images of the NaCl nanoparticle specimen used for the reconstructions provided in subsection \ref{subsec:expdemoNaClparticle}. a) Large field of view of the grid, demonstrating the range of particle size available. b) Closer view of the nanocrystal employed in figure \ref{fig:realtimeprocessNaClparticle}, with indication of lateral size.}
            \label{fig:largefieldofviewNaCl}
        \end{figure}
        
        The TEM grid employed to acquire the data presented in subsection \ref{subsec:expdemoNaClparticle} was prepared using chemical grade NaCl salt (Sigma-Aldrich), with a purity of 99\% and available in powder form. As a first step, an isopropanol suspension was made, thus without relying on an aqueous solvent. This avoids dissolving the native nanoparticles, which would otherwise imply a recrystallization on the support and impede control over size and morphology.
        
        After a sonication step, limiting unwanted aggregations, a volume of 3 $\mu$L was dropcast on an ultrathin (3 nm) film of amorphous carbon, fixed on a standard copper grid (300 mesh). This support was prepared in-laboratory. The ensemble was then left to dry under ambient conditions, finally resulting in isolated NaCl nanocrystals with a broad size distribution. This is illustrated in figure \ref{fig:largefieldofviewNaCl}.a, where an example STEM image is provided, showing a large field of view. The thinnest available particles were preferred for data acquisition.
        
        The site used for later ptychographic treatment, with results given in figure \ref{fig:realtimeprocessNaClparticle}, is depicted in fig. \ref{fig:largefieldofviewNaCl}.b, and consists of a square nanoparticle. While the exact thickness of the illuminated area was not accurately estimated in this work, it is likely comparable to the lateral size of the object, found above 200 nm as indicated over the micrograph.
        
        In parallel, the NaCl bulk specimen employed in subsection \ref{subsec:expdemoNaClbulk} was produced through standard focused ion beam (FIB) milling techniques, from an initial monocrystalline block. This includes lift-off, two-ended fixation on a specimen support and additional FIB-based thinning. As such, controlled reduction of the thickness, down to a nominal value of 30 nm, could be ensured while conserving the crystalline structure, and thus obtaining an appropriate object for ptychographic imaging.

\section*{Funding and acknowledgments}
    
    \textbf{H.L.L.R.} and \textbf{J.V.} acknowledge funding from the Horizon 2020 research and innovation programme (European Union), under grant agreement No. 101017720 (FET-Proactive EBEAM). \textbf{H.L.L.R.}, \textbf{A.A.} and \textbf{J.V.} acknowledge funding from the Horizon Europe programme (European Union) under grant agreement No. 101094299 (Impress). Views and opinions expressed are however those of the authors only and do not necessarily reflect those of the European Union or the European Research Executive Agency (REA). Neither the European Union nor the granting authority can be held responsible for them.
    
    \textbf{T.C.} and \textbf{J.V.} acknowledge further funding from the Flemish government (iBOF project PERsist), and from the Research Foundation - Flanders (FWO, Belgium) under granted project No. G013122N.
    
    Finally, prof. Nini Pryds (Technical University of Denmark) and his group are gratefully acknowledged for providing the SrTiO$_{\text{3}}$ specimen, presented in subsection \ref{subsec:expdemoSTO}.

\section*{Author contribution statement}
    
    \textbf{H.L.L.R.}: concept, theory, methodology, simulation, software, data analysis, writing, supervision. \textbf{A.A.}: methodology, experiment, data analysis, software, draft review. \textbf{T.C.}: methodology, specimen preparation, experiment, simulation, draft review. \textbf{J.V.}: concept, draft review, supervision, funding acquisition.
    
    \textbf{H.L.L.R.} and \textbf{J.V.} are the stated inventors in an international patent application for the GPRI framework \cite{LalandecRobert2025a}, introduced in this publication.

\section*{Declaration of conflicts of interest}
    
    The authors declare no conflicts of interest.

\bibliography{Sources_HLLR.bib}

\end{document}